\documentclass[11pt]{article}
\pdfoutput=1

\usepackage{jheppub} 

\usepackage{amsmath,amssymb,epsfig,amsfonts}
\usepackage{graphicx,subfigure}
\usepackage{epsf}
\usepackage{makeidx}
\usepackage[all]{xy}
\usepackage{slashed}
\usepackage{verbatim} 
\usepackage{float}
\restylefloat{table}
\usepackage[all]{xy}
\usepackage{pdflscape}
\usepackage{tikz}
\usepackage{array}
\usepackage{youngtab}
\usepackage{multirow}
\usepackage{color}
\usepackage{ulem}
\usepackage{cleveref}

\usepackage{tensor}

\makeatletter

\newcommand{\be}{\begin{equation}}
\newcommand{\ee}{\end{equation}}
\newcommand{\ba}{\begin{aligned}}
\newcommand{\ea}{\end{aligned}}

\newcommand{\nn}{\nonumber}

\newcommand{\dd}{\mathrm{d}}
\newcommand{\me}{\mathrm{e}}
\newcommand{\ii}{\mathrm{i}}

\newcommand{\vol}{\mathrm{vol}}

\newlength{\sswidth}

\newcommand{\bea}{\begin{eqnarray}}
\newcommand{\eea}{\end{eqnarray}}

\def\unit{{1\kern-.65ex {\rm l}}}
\def\1{{1\kern-.65ex {\rm l}}}

\newcount\hour \newcount\minute
\hour=\time \divide \hour by 60
\minute=\time
\def\now{%
\ifnum \hour<13
  \ifnum \hour=0 \advance \hour by 12 \number\hour:\else \number\hour:\fi%
     \ifnum \minute<10 0\fi%
     \number\minute%
\ A.M.%
\else \advance \hour by -12 \number\hour:%
  \ifnum \minute<10 0\fi%
  \number\minute%
  \ P.M.%
\fi%
}

\makeatother

\begin{document}

\baselineskip=18pt  
\numberwithin{equation}{section}  
\allowdisplaybreaks  

\begin{titlepage}
\vspace*{-1cm}
\begin{center}
 {\Large\bf ${\cal N}=(2,2)$ AdS$_3$ from D3-branes wrapped on Riemann surfaces}
\end{center}
\vspace*{-.5cm}
%

\begin{center}
{\small
Christopher Couzens$^{a}$\footnote{c.a.couzens@uu.nl},  Niall T. Macpherson$^{b}$\footnote{ntmacpher@gmail.com}, 	Achilleas Passias$^{c, d}$\footnote{achilleas.passias@phys.ens.fr}\\[4mm]}
\vspace{5mm}
$a$: Institute for Theoretical Physics, Utrecht University
Princetonplein 5, 3584 CC Utrecht, The Netherlands\\
\vskip 3mm
$b$: Department of Physics, University of Oviedo, Avda. Federico Garcia Lorca s/n, 33007 Oviedo,
Spain\\
\vskip 3mm
$c$: Department of Nuclear and Particle Physics, 
Faculty of Physics, National and Kapodistrian University of Athens, 
Athens 15784, Greece \\
\vskip 3mm
$d$: Laboratoire de Physique de l’Ecole normale supérieure, ENS, Université PSL, CNRS, Sorbonne Université, Université de Paris, F-75005 Paris, France\\

\vspace*{.5cm}
\end{center}

\abstract
\noindent 
We construct $\mathcal{N}=(2,2)$ supersymmetric AdS$_3$ solutions  of type IIB supergravity, dual to twisted compactifications of 4d $\mathcal{N}=4$ super-Yang--Mills on Riemann surfaces. We consider both theories with a regular topological twist, and a twist involving the isometry group of the Riemann surface. These solutions are interpreted as the near-horizon of black strings asymptoting to AdS$_5\times \text{S}^5$. As evidence for the proposed duality we compute the central charge of the gravity solutions and show that it agrees with the field theory result. 

\end{titlepage}

\tableofcontents


\section{Introduction}

Since the seminal work of \cite{Maldacena:2000mw}, studying the low energy dynamics of branes wrapped on compact manifolds has been a successful research direction. From the field-theoretic point of view one may generate families of lower-dimensional field theories from a higher-dimensional parent theory living on the worldvolume of the branes in flat spacetime. By placing the branes on a non-trivially curved manifold and requiring the preservation of some supersymmetry, one must perform a twist of the field theory. 
Geometrically, this is realised by studying the embedding of a calibrated cycle, on which the brane is wrapped, into a special holonomy manifold. The different ways of embedding the compactification manifold give rise to the different families of lower-dimensional field theories. 

For a large enough number of branes, and small enough curvatures, one may use the holographic correspondence. Since the original work of \cite{Maldacena:2000mw}, which studied the holographic duals of 4d $\mathcal{N}=4$ super-Yang--Mills (SYM) and the 6d $\mathcal{N}=(2,0)$ theory compactified on a Riemann surface, there have been many generalisations. The theories of class $\mathcal{S}$ study the twisted compactification of the 6d $(2,0)$ SCFT living on M5-branes on a punctured Riemann surface \cite{Gaiotto:2009gz} preserving $\mathcal{N}=2$ supersymmetry in four dimensions. These were later extended to theories preserving $\mathcal{N}=1$ supersymmetry in four dimensions in \cite{Bah:2015fwa,Bah:2011je,Bah:2012dg,Bah:2013qya} and the holographic dual of Argyres--Douglas theories in \cite{Bah:2021mzw,Bah:2021hei}. Theories of class $\mathcal{R}$ study the compactification of the 6d $(2,0)$ $A_{N-1}$ theory on hyperbolic three-manifolds, \cite{Bobev:2019zmz}. Whilst the theories in \cite{Bah:2018lyv} arise from compactifying the D4-D8 bound state on a Riemann surface. In addition, the work of \cite{Bobev:2019ore} studies various compactifications of branes on Riemann surfaces with punctures using gauged supergravity, and \cite{Ferrero:2020laf,Ferrero:2021wvk,Boido:2021szx,Hosseini:2021fge} consider field theories wrapped on spindles. 

In this work we will restrict to studying the holographic dual of a stack of D3-branes compactified on a Riemann surface with a twist preserving $\mathcal{N}=(2,2)$ supersymmetry and flowing to a two-dimensional SCFT.\footnote{Recently constraints on the elliptic genera of 2d $\mathcal{N}=(2,2)$ SCFTs admitting holographic duals have been investigated in \cite{Belin:2020nmp,Belin:2019jqz,Belin:2019rba}. Though we will not compute the elliptic genus of the field theory duals of our solutions it would be interesting to see where these theories lie in the landscape mapped out there.}
The UV theory is 4d $\mathcal{N}=4$ SYM which flows to the 2d SCFT after being placed on the Riemann surface. Via the AdS/CFT correspondence the two theories, the UV and IR, admit AdS duals. For the UV theory this is of course AdS$_5\times \text{S}^5$ whilst for the IR one obtains the AdS$_3$ solutions which form the content of this work. The full flow solution is a five-dimensional asymptotically AdS$_5$ black string with AdS$_3\times \Sigma_2$ near-horizon geometry.\footnote{Static black string solutions in 5d STU supergravity have been found in \cite{Azzola:2018sld} for example.} 

The  microscopic origin of the Bekenstein--Hawking entropy of black objects is one of the fundamental problems in theoretical physics. Certain thermodynamic quantities of black objects may be computed in gravity in either the asymptotic or near-horizon limit without loss of information. One such quantity is the Bekenstein--Hawking entropy. The fact that one only needs the near-horizon to compute the Bekenstein--Hawking entropy and not the full black string solution is a considerable simplification to the problem. One can then identify the microstates of the black string as the SCFT dual to the AdS$_3$ near-horizon theory. In this paper we will take this latter approach and interest ourselves in the near-horizon of black strings preserving $\mathcal{N}=(2,2)$ supersymmetry and not the full black string solution.

Solutions of this type were classified in \cite{Couzens:2017nnr}, whilst AdS$_3$ solutions preserving other amounts of supersymmetry have been studied in \cite{Martelli:2003ki,Tsimpis:2005kj,Kim:2005ez,Kim:2007hv,Figueras:2007cn,Donos:2008hd,Colgain:2010wb,DHoker:2008lup,Estes:2012vm,Bachas:2013vza,Jeong:2014iva,Lozano:2015bra,Kelekci:2016uqv,Couzens:2017way,Eberhardt:2017uup,Dibitetto:2018iar,Dibitetto:2018ftj,Macpherson:2018mif,Legramandi:2019xqd,Lozano:2019emq,Lozano:2019jza,Lozano:2019zvg,Lozano:2019ywa,Couzens:2019mkh,Couzens:2019iog,Passias:2019rga,Filippas:2019ihy,Speziali:2019uzn,Lozano:2020bxo,Farakos:2020phe,Couzens:2020aat,Rigatos:2020igd,Faedo:2020nol,Dibitetto:2020bsh,Filippas:2020qku,Passias:2020ubv,Faedo:2020lyw,Eloy:2020uix,Legramandi:2020txf,Zacarias:2021pfz,Emelin:2021gzx,Couzens:2019wls}. The conditions for preserving supersymmetry and satisfying the equations of motion, which we review in section \ref{sec:22solutions}, are very similar to the conditions for other AdS solutions which are identified as wrapped brane solutions. In particular the form of the solutions are similar to the AdS$_5$ solutions arising from M5-branes on a Riemann surface \cite{Bah:2013qya} and AdS$_4$ solutions in massive type IIA obtained from the D4-D8 bound state on a Riemann surface, \cite{Passias:2018zlm,Bah:2018lyv}. We will reduce the conditions of \cite{Couzens:2017nnr} to a Monge--Amper\'e like equation for a potential. We will present two distinct classes of solutions. The first has a constant curvature metric on the Riemann surface and we find that the solution is a deformation of the Maldacena--Nu$\tilde{\text{n}}$ez solution \cite{Maldacena:2000mw}. The benefit of our formulation is that the inclusion of punctures has a natural interpretation in this description: they are interpreted as sources to the Monge--Amper\'e equation. The second class of solutions we study has a non-constant curvature metric on the Riemann surface wrapped by the D3-branes. Topologically the space is a disk with a $\mathbb{Z}_{k}$ orbifold singularity at the centre and smeared D3-branes on the boundary.

The paper is organised as follows. In section \ref{sec:22solutions} we will review the construction of AdS$_3$ solutions in type IIB preserving $\mathcal{N}=(2,2)$ supersymmetry before reducing the conditions under the assumption of the existence of a flavour symmetry. We next consider solutions where the branes wrap a constant curvature Riemann surface in section \ref{sec:H2}. We begin by reviewing the field theory in section \ref{sec:SCFT} before constructing a simple holographic dual in section \ref{sec:explicit} and analysing the regularity and central charge of the solution in section \ref{sec:reg}. We generalise this solution further in section \ref{sec:gen} before computing additional observables of the theory and comparing to the field theory results. 
In section \ref{sec:disk} we consider the holographic duals of 4d $\mathcal{N}=4$ SYM on a topological disk equipped with a non-constant curvature Riemann surface, analysing the solutions in detail. 
Some technical material is relegated to two appendices.



\section{$\mathcal{N}=(2,2)$ AdS$_3$ solutions}\label{sec:22solutions}

We now turn our attention to constructing the supergravity dual of the 2d $\mathcal{N}=(2,2)$ SCFTs discussed in the introduction. We will reduce $\mathcal{N}=4$ SYM on a Riemann surface, breaking the SU$(4)_{\rm R}$ R-symmetry to its U$(1)^3$ Cartan in order to perform a topological twist. As such, the resultant two-dimensional field theory should admit three U$(1)$ symmetries: two will furnish the U$(1)_L\times$U$(1)_{R}$ R-symmetry of the $\mathcal{N}=(2,2)$ superconformal algebra whilst the third will be a flavour symmetry of the theory. Given these considerations, the metric of the gravity solution should take the form
\be
\text{AdS}_{3} \times \left( \Sigma_2\times \text{S}^{1}_{R}\times \text{S}^{1}_{L}\times \text{S}^{1}\times I_1\times I_2 \right)\, ,\label{idealisedsol}
\ee
where $I_{1}$, $I_{2}$ are two line intervals. Solutions of this form, supported by five-form flux were classified in \cite{Couzens:2017nnr}.\footnote{For convenience we make a few redefinitions of the results in \cite{Couzens:2017nnr}, in particular we redefine the warp factor via $\me^{-4 \Delta}y= \sin^2 \zeta$.} The metric is given by
\be 
\frac{1}{L^2}\dd s^2=\frac{\sqrt{y}}{\sin \zeta}\left[ \dd s^2(\text{AdS}_{3})+\dd s^{2}(X_{7})\right]\, ,
\ee
with AdS$_3$ having unit radius and $L$ a constant.  
Supersymmetry fixes the internal metric to take the form 
\be \label{eq:7manifold}
\dd s^2 (X_7)= \cos^{2}\zeta (\dd \psi_1+ \sigma)^2+ \sin^{2}\zeta \dd \psi_2^2+\frac{\sin^2\zeta}{4 y^2 \cos^2\zeta }\dd y^2+ \frac{\sin^2\zeta}{y} g^{(4)}(y, x)_{ij} \dd x^{i} \dd x^{j}\, .
\ee
The coordinate $y$ defines an almost-product structure, implying that the one-form $\sigma$ has ``legs'' only along the four-dimensional base, though it may still depend on $y$ non-trivially. 
The metric $g^{(4)}$ on the base admits an SU$(2)$-structure $(J,\,\Omega)$ satisfying\footnote{One can reduce the number of these conditions further by reinterpreting the conditions in terms of derivatives of the determinant of the metric, see \cite{Couzens:2017nnr}. This reformulation is reminiscent of the ones appearing for AdS$_5$ solutions in M-theory \cite{Gauntlett:2004zh} and AdS$_4$ solutions in massive type IIA \cite{Passias:2018zlm}. However, for our purposes the most convenient formulation of the conditions are the ones we present here.}
\begin{align}
\partial_{\psi_1}J&=\partial_{\psi_2}J=0\, ,\label{Jpsi}\\
\partial_{y} J&= \frac{1}{2} \dd_4 \sigma\, ,\label{Jy}\\
\dd_4 J&=0\, ,\label{Jd4}
\end{align}
and
\begin{align}
\partial_{\psi_1}\Omega&=-\ii \Omega\, ,\label{Omegapsi1}\\
\partial_{\psi_2}\Omega&=0\, ,\label{Omegapsi2}\\
\partial_{y} \Omega&= \frac{1}{2} \tan^2 \zeta \partial_{y}\left( \log \left(\frac{\sin^2 \zeta}{y}\right) \right)\Omega\, ,\label{Omegay}\\
\dd_4 \Omega&= -\left[ \ii \sigma + \dd_4(\log \cos\zeta) \right] \wedge \Omega\, .\label{Omegad4}
\end{align}
The metric is supported by the five-form flux
\be
\frac{1}{L^4}F_5=(1+\star)\dd \vol(\mathrm{AdS}_3)\wedge F^{(2)}\, ,
\ee
with  
\be
F^{(2)}=y \dd \sigma -2 J_4 - \dd \left( \frac{y}{\sin^2\zeta}(\dd \psi_1+\sigma)\right)\, .
\ee
Supersymmetry implies that all equations of motion are satisfied provided all the torsion conditions given above hold. Contrast this with the $\mathcal{N}=(0,2)$ case \cite{Kim:2005ez}, where one must also impose the Maxwell equation for $F^{(2)}$. 

Consistent with these geometries preserving $\mathcal{N}=(2,2)$ supersymmetry, which has R-symmetry group U(1)$_{L}\times$U(1)$_{R}$, the metric admits two independent Killing vectors: $\partial_{\psi_{
1}}$, $\partial_{\psi_{2}}$. The generator dual to the left-moving R-symmetry is $\partial_{\psi_{1}}-\partial_{\psi_{2}}$ whilst the right moving is dual to $\partial_{\psi_{1}}+\partial_{\psi_{2}}$. Recall that the solution we are looking for should have three U$(1)$ isometries. The general form of the internal metric accounts for two of the three required, and therefore we require that the metric $g^{(4)}$ should admit a single Killing vector.

\subsection{Embedding into the $(0,2)$ classification}\label{sec:(0,2)}

Before we proceed, it is useful to study the embedding of the $\mathcal{N}=(2,2)$ conditions into the classification of $\mathcal{N}=(0,2)$ solutions \cite{Kim:2005ez}. We first want to isolate the right-moving R-symmetry vector field, to do this we make the change of  coordinates
\be
\psi_{1}=\frac{1}{2}(\phi_R+\phi_L)\, , \quad \psi_{2}=\frac{1}{2}(\phi_{R}-\phi_{L})\, .
\ee
A small rearrangement of terms after the coordinate transformation, and identifying the warp factor to be
\be
y\me^{-4 \Delta}=\sin^2\zeta\, ,
\ee
puts the solution into the form
\be
\dd s^2(X_7)=\frac{1}{4}\left[\dd \phi_{R}+(1-2 y \me^{-4 \Delta})\dd \phi_{L}+2(1-y \me^{-4 \Delta})\sigma \right]^2+ \me^{-4 \Delta} \dd s^2(Y_6)\, ,
\ee
where 
\be
\dd s^2 (Y_6)= \frac{1}{4 y (1- y \me^{-4 \Delta})} \dd y^2+ y (1-y \me^{-4 \Delta}) (\dd \phi_{L}+\sigma)^2+ g(y,x)_{ij}\dd x^i \dd x^j\, .
\ee
This is the same form for the metric in the $\mathcal{N}=(0,2)$ classification \cite{Kim:2005ez}.\footnote{We use the conventions in \cite{Couzens:2017nnr} rather than \cite{Kim:2005ez}, the difference is a minor change of orientation.} The one-form dual to the R-symmetry vector is 
\be
\eta = \frac{1}{2}\left[\dd \phi_{R}+(1-2 y \me^{-4 \Delta})\dd \phi_{L}+2(1-y \me^{-4 \Delta})\sigma \right]\, ,
\ee 
and the K\"ahler base $Y_6$ is defined above. The SU$(3)$-structure forms, in terms of the SU$(2)$-structure ones are given by
\be
J_6 = \frac{1}{2} \dd y \wedge ( \dd \phi_{L}+\sigma)+ J\, , \quad \Omega_6=\frac{1}{2 \sqrt{y(1- y \me^{-4 \Delta})}}\left[\dd y + 2 \ii  y (1- y \me^{-4 \Delta})(\dd \phi_{L}+ \sigma)\right] \wedge \Omega\, ,
\ee
and it is simple to see that they satisfy the $\mathcal{N}=(0,2)$ conditions,
\begin{align}
\dd J_6=0\, , \quad \dd \Omega_6=-2 \ii \eta \wedge \Omega_6\, ,
\end{align}
provided the $\mathcal{N}=(2,2)$ torsion conditions \eqref{Jpsi}-\eqref{Omegad4} hold.
Moreover, since $\mathcal{N}=(2,2)$ supersymmetry imposes the Maxwell equation for $F^{(2)}$ it follows that the $\mathcal{N}=(0,2)$ master equation is also satisfied. 

\subsection{Canonical form of solutions with $\mathbb{T}^3$ fibration}\label{sec:generalform}
In appendix \ref{sec:derivation} we have reduced the torsion conditions \eqref{Jpsi}-\eqref{Omegad4} under the assumption that $g^{(4)}$ contains an additional flavour U(1) direction. The solution is determined by a potential $D$ satisfying a Mong\'e--Ampere like equation \eqref{MAeq}, which is a generalisation of the Toda equation. Specifically this takes the form
\be
\square D=16 y^2 \Big( \partial_{y}^2 D \partial_{\Theta} D-(\partial_{y}\partial_{\Theta}D)^2\Big) \me^{\partial_{y}D}\, ,\label{MAeq}
\ee
with $\square\equiv \partial_{X_{1}}^2+\partial_{X_2}^2$. Once a solution to \eqref{MAeq} has been provided the metric is given by
\begin{align}
\frac{1}{L^2}\dd s^2&=\sqrt{\frac{y g}{h}} \left[ \dd s^2 (\text{AdS}_3) +\frac{h}{g}\dd \psi_{2}^2 +\frac{h\me^{2 A}}{y g}(\dd X_1^2+\dd X_2^2)+\frac{h}{yg}\dd s^{2}(\mathcal{M}_4)\right]\, ,\nonumber\\[2mm]
\dd s^2 (\mathcal{M}_{4})&= \frac{1}{4} g_{ij}\dd u^i\dd u^j+h^{ij} \eta_{i}\eta_{j}\, , \quad \me^{2A}= 4y^2\me^{\partial_y D}g\, ,\nonumber\\[2mm]
g_{ij}&\equiv-\partial_{i}\partial_{j}D\, ,\quad h_{ij}\equiv-\partial_{i}\partial_{j}\left(D+y(\log y-1)\right)\, ,\nonumber\\[2mm]
\eta_{1}&\equiv\dd\psi_{1}+\star_{2}\dd_2 (\partial_{y}D)\, ,\quad \eta_{2}\equiv\frac{1}{2}\left(\dd \phi+ \star_{2}\dd_2 (\partial_{\Theta} D) \right)\, ,\label{eq:generalform}
\end{align}
with 
\be
u^{i}=\{y,\Theta\}\, ,\quad g\equiv\det(g_{ij})\, ,\quad h\equiv\det(h_{ij})\,.
\ee
Dualising along the non-R-symmetry directions to type IIA and then uplifting to M-theory one will obtain an AdS$_3$ solution in the class discussed in \cite{Bah:2015nva}. It would be interesting to study the solutions discussed in the following sections from an M-theory perspective. 

The form of the metric suggests that one should look for solutions with D3-branes wrapped on the $(X_1,\,X_2)$ directions; we will do just that in section \ref{sec:H2}. However, we should stress that these directions can lie in the co-dimensions of the wrapped D3-branes as we shall find in section \ref{sec:disk}. Our  focus in this work  are  cases where $(X_1,\,X_2)$ span a Riemann surface of constant curvature.  Going away from the constant curvature metric and including punctures on the Riemann surface is beyond the scope of this work, however the generalisation of our ansatz in this direction is obvious, though technically challenging.


\section{D3-branes wrapping a constant curvature Riemann surface}\label{sec:H2}

In this section we will consider the compactification of 4d $\mathcal{N}=4$ SYM on a constant curvature Riemann surface $\Sigma_\mathfrak{g}$, of genus $\mathfrak{g}$, preserving $\mathcal{N}=(2,2)$ supersymmetry. We begin in section \ref{sec:SCFT} by reviewing the construction of the 2d SCFT arising from the compactification. In particular, we will compute the central charge and the R-charges of baryonic operators with which we may compare to a dual gravity computation. Our gravity analysis begins with the reduction of the general $\mathcal{N}=(2,2)$ supersymmetry conditions of \eqref{sec:generalform} for a constant curvature Riemann surface. We find that there is no solution for $\mathfrak{g}=0$, whilst the solution for $\mathfrak{g}=1$ has an enhancement of supersymmetry to $\mathcal{N}=(4,4)$ and is the double T-dual of the D1-D5 system. For $\mathfrak{g} > 1$ we find a family of solutions which generalise the Maldacena--Nu$\tilde{\text{n}}$ez solution \cite{Maldacena:2000mw}, breaking the SU$(2)$ symmetry of their solution to U$(1)$. We present the regularity analysis of these solutions before computing their central charge and R-charges, which match with the field theory analysis in section \ref{sec:SCFT}.

\subsection{Twisted D3-branes}\label{sec:SCFT}
 Let us consider compactifications of 4d $\mathcal{N}=4$ SYM on a Riemann surface. As is well known when the Riemann surface has non-trivial curvature it is necessary to perform a topological twist of the theory in order to preserve supersymmetry.\footnote{Recently, compactifications which twist with the isometry of the compactification space have been investigated, see for example \cite{Bah:2019rgq}. } Let us begin by defining our conventions for 4d $\mathcal{N}=4$ SYM. For the moment we will use $\mathcal{N}=4$ representations, however as we will explain later, in order to compare to the R-charges computed in gravity it is useful to rewrite the theory in terms of $\mathcal{N}=1$ multiplets. 

As is well known 4d $\mathcal{N}=4$ SYM contains a single $\mathcal{N}=4$ gauge multiplet, which in $\mathcal{N}=1$ language consists of a single $\mathcal{N}=1$ gauge multiplet and three chiral superfields $\Phi_{i}$. The bosonic symmetry group of the theory is 
\be
\text{SO}(1,3)\times \text{SU}(4)_{\rm R}\, ,
\ee
with SU$(4)_{\rm R}$ the R-symmetry of the theory. The 16 supersymmetry transformation parameters (Killing spinors) transform  under the bosonic symmetry group as
\be
\epsilon : ({\bf 2}, {\bf 1}, {\bf 4})\,,\quad \tilde{\epsilon} : ({\bf 1}, {\bf 2}, \bar{{\bf 4}})\, ,
\ee
whilst the gauginos transform in
\be
\Psi : ({\bf 2}, {\bf 1} ,{\bf 4})\, ,\quad \tilde{\Psi} : ({\bf 1}, {\bf 2}, \bar{{\bf 4}})\, .
\ee
We now want to place the theory on a Riemann surface of constant curvature $\Sigma_{\mathfrak{g}}$, such that the Ricci scalar is $R_{\Sigma_{\mathfrak{g}}}=2\kappa$ with $\kappa=\{-1,0,1\}$, for $\Sigma_{\mathfrak{g}>1} = \text{H}^2$, $\Sigma_{\mathfrak{g}=1} =\mathbb{T}^2$, and $\Sigma_{\mathfrak{g}=0} =\text{S}^2$ respectively. For H$^2$ this is meant locally, globally it is a compact quotient H$^2/\Gamma$,  where $\Gamma$ is an element of the Fuschian subgroup of the PSL(2, $\mathbb{R}$) isometry group of H$^2$. The theory then flows to a two-dimensional SCFT in the IR. The presence of the Riemann surface breaks the four-dimensional Lorentz symmetry to
\be
\text{SO}(1,3)\rightarrow \text{SO}(1,1) \times \text{U}(1)_{\Sigma_{\mathfrak{g}}}\, .
\ee
Generically this does not preserve any supersymmetry unless the Riemann surface is the two-torus $\mathbb{T}^2$. Since we are interested in preserving $(2,2)$ supersymmetry in the resultant 2d SCFT we must perform a topological twist. We will twist with the Cartan of the SU$(4)_{\rm R}$ R-symmetry, namely U$(1)^3$. 
Under the Cartan the representations of SU$(4)_{\rm R}$ that we need, decompose as (see for example \cite{Lawrie:2016axq}):
\begin{align}
\text{SU}(4)_{\rm R}&\rightarrow \text{U}(1)_1\times \text{U}(1)_2\times \text{U}(1)_3\nonumber\\[2mm]
{\bf 4}&\rightarrow  {\bf 1}_{\tfrac{1}{2},\tfrac{1}{2},\tfrac{1}{2}}\oplus{\bf 1}_{\tfrac{1}{2},-\tfrac{1}{2},-\tfrac{1}{2}}\oplus{\bf 1}_{-\tfrac{1}{2},\tfrac{1}{2},-\tfrac{1}{2}}\oplus{\bf 1}_{-\tfrac{1}{2},-\tfrac{1}{2},\tfrac{1}{2}}\label{4decomp}\\[2mm]
{\bf 6}&\rightarrow {\bf 1}_{1,0,0}\oplus{\bf 1}_{0,1,0}\oplus{\bf 1}_{0,0,1}\nonumber\, .
\end{align}
We now want to study the possible twists of the theory which preserve $\mathcal{N}=(2,2)$ supersymmetry. Such twists have been classified previously in \cite{Maldacena:2000mw, Benini:2013cda,Amariti:2017cyd,Bobev:2019ore}, however for completeness and to extract out the R-charges, we will perform the analysis here as well. We turn on a background gauge field $A$ along the generator $T$
\be
T=a_1 T_1+a_2 T_2 +a_3 T_3\, ,
\ee
with $T_i$ the generator of the $i$'th U$(1)$ in the U$(1)^3$ Cartan. 
Supersymmetry of the theory is preserved subject to a Killing spinor equation being satisfied. By turning on a background gauge field, the Killing spinor equation on the Riemann surface is given by
\be
\left(\partial_{\mu}-\frac{\ii}{2} s \omega_{\mu}-\ii  A_{\mu}\right)\epsilon=0\,.
\ee
The gauge field $A$ has field-strength $F= -\kappa T \dd \omega$, for $\kappa\neq0$, with $\omega$ the spin connection on the Riemann surface, whilst for $\kappa=0$ we have $F= - T \frac{2\pi}{\vol(\mathbb{T}^2)}\dd \vol(\mathbb{T}^2)$. The constant $s$ is $\pm1$ for positive/negative chirality spinors respectively. In order for supersymmetry to be preserved we must tune the parameters $a_{i}$ such that the Killing spinor equation admits constant spinor solutions. This requires the contribution from the spin connection to be cancelled by the background gauge field.
 
Using the decomposition in \eqref{4decomp} and denoting the spinors as $\epsilon_{\pm}^{m}$, $m \in \{1,2,3,4\}$ we see that constant spinors satisfy the Killing spinor equation provided
\begin{align}
\mp \frac{\kappa}{2}=a_1+a_2+a_3\, ,\quad \mp\frac{\kappa}{2}= a_1-a_2-a_3\, ,\quad \mp\frac{\kappa}{2}= -a_1+a_2-a_3\, ,\quad \mp\frac{\kappa}{2}= -a_1-a_2+a_3\, ,
\end{align}
for each of the spinors $\epsilon_{\pm}^{m}$ respectively.

$\mathcal{N}=(2,2)$ supersymmetry is preserved if the $a_{i}$ are chosen such that two of the conditions above are satisfied, one for a positive chirality spinor and one for a negative chirality spinor. Without loss of generality, we can choose these spinors to be $\epsilon_{+}^1$ and $\epsilon_{-}^{2}$, which implies that the $a_{i}$'s satisfy
\be\label{eq:aFT}
a_{1}=0\, ,\quad a_{2}+a_{3}=-\kappa\, .
\ee
Recall that a 2d $\mathcal{N}=(2,2)$ SCFT admits a U$(1)_{L}\times$U$(1)_{R}$ R-symmetry. We may identify the right-moving R-symmetry by requiring that the positive chirality spinor has R-charge 1 and the negative chirality spinor has vanishing R-charge under it. Conversely under the left-moving R-symmetry the positive chirality spinor has vanishing R-charge whilst the negative chirality spinor has R-charge 1. 
The most general right-moving U$(1)$ R-symmetry with these properties is given by\footnote{These choices of R-symmetry generalise those taken in \cite{Maldacena:2000mw}.}
\be
{\rm R}_{R}=T_{1}+ \cos^2 \alpha_{R} T_{2}+\sin^2 \alpha_{R} T_3\, ,\label{FTRR}
\ee
whilst for the left-moving R-symmetry we have
\be
{\rm R}_{L}=T_{1}- \cos^2 \alpha_{L} T_2-\sin^2\alpha_{L} T_3\, .
\ee
Note in addition that
\be
T_{\rm F}=T_2-T_3\, ,
\ee
parametrises a flavour symmetry of the theory. For a theory which is conformal at the IR fixed point it is necessary that the  `t Hooft anomaly\footnote{The indices $I,J$ run over all U$(1)$'s of the theory. Here these are a flavour symmetry, and a left- and right-moving R-symmetry. The $Q$'s are the charges of the Weyl fermion under the generator of the symmetry and $\gamma^3$ is the chirality matrix in 2d.} 
\be
k_{IJ}=\sum_{\text{Weyl fermions} }\gamma_3 Q_{I} Q_{J}\, ,
\ee
of a left- and right-moving current vanishes \cite{Benini:2013cda}. In addition at the fixed point the `t Hooft anomaly for a flavour and R-symmetry must vanish, therefore altogether we must impose 
\be 
k_{R{\rm F}}=k_{L{\rm F}}=k_{LR}=0\, .\label{contraintfieldtheory}
\ee
Note that the first two are the conditions imposed by $c$-extremization, whilst the latter is an additional constraint which imposes $\mathcal{N}=(2,2)$ supersymmetry. As an aside, if one naively performs $c$-extremization preserving $\mathcal{N}=(0,2)$ supersymmetry and then tunes the parameters to preserve $\mathcal{N}=(2,2)$ supersymmetry as in \eqref{eq:aFT} one finds the non-sensical result that the central charge vanishes. By imposing the additional constraint that $k_{LR}=0$ we alleviate this problem.

In order to evaluate the `t Hooft anomalies we need to know the multiplicities of the 2d fermions. This can be computed by using the Riemann--Roch index theorem. The difference between the number of zero modes on $\Sigma_{\mathfrak{g}}$ is given by \cite{Maldacena:2000mw}
\be
n^{+}-n^{-}= \frac{1}{2\pi} \int_{\Sigma_{\mathfrak{g}}} F= 2 T(\mathfrak{g}-1)\, .
\ee
The constraints from \eqref{contraintfieldtheory} are satisfied provided
\be
a_{2}=a_{3}=-\frac{\kappa}{2}\, ,
\ee
which gives the following `t Hooft anomalies for the left- and right-moving R-symmetries 
\be
k_{RR}=k_{LL}=N^2 (\mathfrak{g}-1)\, .
\ee
Consequently the central charges are given by
\be
c_{L}=c_{R}=3 N^2 (\mathfrak{g}-1)\, ,
\ee
where we have used $c_{R}=3 k_{RR}$ and $c_{L}=3 k_{LL}$.
Note that the central charge is only well-defined for $\mathfrak{g}>1$, the compactification on the two-sphere with this twist does not give rise to a well defined IR SCFT. Note in particular that the parameters of the R-symmetry drop out of the central charge. To see the contribution of these parameters we should compute the R-charges of the reduced scalar fields. Recall that the scalars of 4d $\mathcal{N}=4$ SYM transform in the $\bf 6$ of the SU$(4)_{\rm R}$ R-symmetry. From \eqref{4decomp} we have that the $i$'th $\mathcal{N}=1$ chiral multiplet may be arranged to have charge $1$ under the $i$'th U$(1)$ and uncharged under the others. It then follows from \eqref{FTRR} that the (right-moving) R-charges are
\be
{\rm R}[\Phi_1]=1\, ,\quad {\rm R}[\Phi_{2}]=\cos^2\alpha_{R}\, ,\quad {\rm R}[\Phi_{3}]=\sin^2 \alpha_{R}\, .\label{FTRcharges}
\ee
We will see later that these are indeed reproduced by the gravity solution of section \ref{sec:gen-observables}.

\subsection{A simple ansatz for constant curvature Riemann surfaces}\label{sec:explicit} 
In this section we consider a simple ansatz for D3-branes wrapping Riemann surfaces of constant curvature. As we shall see only the cases of $\mathbb{T}^2$ and H$^2$ give rise to sensible holographic duals, the former reproduces the D3-D3 near-horizon while the latter generalises the ${\cal N}=(2,2)$ solution of \cite{Maldacena:2000mw}. Our starting point is the system presented in section \ref{sec:generalform}.\\
~~\\
In order to make the comparison with the field theory more manifest it is useful to first make the change of coordinates
\be
y= m^{+} t^{+}+m^{-} t^{-}\, ,\quad \Theta= t^{+}-t^{-}\, ,
\ee
where $m^{\pm}$ are constrained to satisfy
\be 
m^{+}+m^{-}=1\, .
\ee
This change of coordinates puts the defining equation into the form
\be
\square D=16 (m^{+}t^{+}+m^{-}t^{-})^2 \Big(\partial_{t^+}^2D \partial_{t^-}^2 D-(\partial_{t^+}\partial_{t^-}D)^2 \Big)\me^{(\partial_{t^+}+\partial_{t^{-}})D}\, .\label{MAeqtcoord}
\ee
We will make the assumption that the Riemann surface is of constant curvature, as such we assume that the warp factor $\me^{2A}$ appearing in front of the Riemann surface takes the factorised form
\be
\me^{2A}= f(t^{+},t^{-}) \me^{2 A_{0}(X_1,X_{2})}\, ,
\ee
where the potential $A_{0}$ satisfies\footnote{In order to generalise to include punctures on the Riemann surface one should add source terms to the right-hand side of equation \eqref{eq:lapforRS}.}
\be
\square A_{0} =-\kappa \me^{2A_0}\, .\label{eq:lapforRS}
\ee
Given the definition of $e^{2A}$, after substituting this ansatz into \eqref{MAeq}, the potential $D$ is required to take the form
\be
D=\mu_{g}^2\tilde{A}_{0} +I(t^{+},t^{-})\, ,
\ee
where
\be
\mu_{g}^2 \equiv 2 (a^{+} t^{+}+ a^{-} t^{-}) + c \, ,
\ee
with $c$ an integration constant and $a^{\pm}$ constants satisfying
\be
a^{+}+a^{-}=\kappa\, .
\ee
In addition we have defined the scalar $\tilde{A}_{0}$ which satisfies
\be
\square \tilde{A}_{0}=-\me^{2 A_{0}}\, ,
\ee
and has solution 
\be
\tilde{A}_{0}=\begin{cases}
-\frac{1}{4}(X_{1}^2+X_{2}^2)&\text{for }\kappa=0\, ,\\[2mm]
\kappa A_{0}& \text{for }\kappa \neq 0\, .
\end{cases}
\ee
With these definitions equation \eqref{MAeqtcoord} reduces to 
\be
-\mu_{g}^2=16 (m^{+}t^{+}+m^{-}t^{-})^2 \me^{(\partial _{t^+}+\partial_{t^{-}})I(t^{+},t^{-})}
\Big(\partial_{t^+}^{2}I \partial_{t^{-}}^2 I-(\partial_{t^+}\partial_{t^-}I)^2 \Big)\, ,
\ee
which we must solve for $I$. Notice that this equation is left invariant when $I$ is shifted by any linear function of $(t^+-t^-)$, so we can without loss of generality  set such terms to zero. 
We will take a similar ansatz to that used in \cite{Bah:2018lyv}, namely
\begin{align}
I=&- \mu_0^2(\log \mu_0^2-1)- \frac{1}{c^{+}}\mu_{+}^{2}(\log \mu_{+}^2-1)-\frac{1}{c^{-}}\mu_{-}^2 (\log \mu_{-}^2 -1)+2 \nu (t^++ t^-)\, ,
\end{align}
where we define 
\be
\mu^{2}_{\pm}=b^{\pm}-c^{\pm} t^{\pm}\,,\quad \mu_0^2= m^+ t^++ m^- t^-= y \, .
\ee
In general the constants are fixed as
\be
a^{\pm}=\frac{1}{2}\kappa\, ,\quad m^{\pm}= \frac{1\pm \epsilon}{2}\, ,\quad c^{\pm}\neq 0\, ,
\ee
with additional constraints that depend on the curvature of the Riemann surface, namely
\begin{align}\label{eq:topologydepconds}
 \kappa=0:&~~~~ \epsilon=\pm 1\, ,~~~~ c=-16 b^{\pm}c^{\mp} e^{4\nu}\, ,\\[2mm]
\kappa\neq0:&~~~~\epsilon= \frac{\kappa (b^{-}c^{+}-b^{+}c^{-})\pm \sqrt{c^{+}c^{-}(c^2c^+c^- -4 b^{-}b^{+})}}{\kappa(b^{-}c^{+}+b^{+}c^{-})+ c\,c^{-}c^{+}}\,,~~~~\me^{4\nu}= -\frac{\kappa}{4 c^{+}c^{-}(1-\epsilon^2)} 
\nn\, .
\end{align}
Notice that we have
\be\label{eq:emedding}
c^+c^-\mu_0^2+ m^+ \mu^2_++m^- \mu^2_-=c^+ b^- m^-+c^- b^+ m^+\, ,
\ee
so that when $m^{\pm}\neq 0$ and likewise the right-hand side of this expression, $(\mu_0,\,\mu_+,\,\mu_-)$  define an embedding of a surface into a  three-dimensional ambient space. The metric of the constant curvature solutions take a simple form in terms of these coordinates, we have in general that
\begin{align}
\frac{1}{L^2}\dd s^2&=\sqrt{\Lambda}\left[\dd s^2(\text{AdS}_3)+  4 e^{4\nu}c^+c^-\me^{2 A_{0}} (\dd X_1^2+\dd X_2^2) \right]+\frac{1}{\sqrt{\Lambda}}\dd s^2 (\mathcal{M}_{5})\, ,\label{eq:metsol}\\[2mm]
\dd s^2 (\mathcal{M}_{5})&= \dd\mu_0^2+\frac{1}{c^+}\dd\mu_+^2+\frac{1}{c^-}\dd\mu_-^2+ \frac{1}{c^+}\mu_+^2\eta_+^2+\frac{1}{c^-}\mu_-^2\eta_-^2+\mu_0^2\dd\psi_2^2\, ,\nn
\end{align}
where we define the warp factor
\be
\Lambda \equiv \mu_0^2+ \frac{(m^+)^2}{c^+}\mu^2_++\frac{(m^-)^2}{c^-}\mu_-^2\, ,
\ee
and the fibered terms 
\be
\eta_{\pm}\equiv\frac{1}{2}\left[(1\pm \epsilon)\dd\psi_1\pm \dd\phi+V\right]\,,\quad V=\kappa\left(\partial_{X_2}\tilde{A}_0\,   \dd X_1-\partial_{X_1}\tilde{A}_0\, \dd X_2\right)\, .
\ee
Clearly, positivity of $\mathcal{M}_{5}$ demands $c^{\pm}>0$ and this manifold  has the appearance of a U(1)$^3$ preserving deformed five-sphere. However, as we will see shortly, the actual topology of this space depends on that of the Riemann surface. We shall study the distinct solutions in more detail in the  next section.



\subsection{Analysis of solutions}\label{sec:reg}

We saw in the previous section that the solution has the appearance of a U$(1)^3$ preserving deformed five-sphere, fibered over a Riemann surface, subject to the embedding condition \eqref{eq:emedding}. In this section we will study the regularity of the solutions for all choices of Riemann surface.

\subsubsection{$\mathbb{T}^2$ case}\label{sec:T2}
We begin by considering the case $\kappa=0$, taking the solution 
\be
\epsilon=1 \, , \quad c=-16e^{4\nu} b^{+}c^{-} \,,
\ee
of \eqref{eq:topologydepconds} without loss of generality. The condition \eqref{eq:emedding} then informs us that we should fix
\be
c^+\mu_0^2=b^+\sin^2\alpha\, , \quad \mu_+^2= b^+ \cos^2\alpha\, ,
\ee
and one finds that $\eta_{\pm}$ are merely two independent linear combinations of $(\psi_1,\,\phi)$,  that we shall identify as $\eta_{\pm}=d\tau_{\pm}$ so that
\be
\dd s^2 (\mathcal{M}_{5})= \frac{b^+}{c^+}\left(\dd\alpha^2+ \cos^2\alpha \dd\tau_+^2+\sin^2\alpha \dd\psi_2^2\right)+\frac{1}{c^-}\left(\dd\mu_-^2+\mu_-^2 \dd\tau^2_-\right)\,,\quad\Lambda= \frac{b^+}{c^+}\, ,
\ee
in other words the warp factor  is constant and the internal five-manifold is  S$^3\times \mathbb{R}^2$. Locally there  is of course no difference between this and S$^3\times \mathbb{T}^2$, making the entire space locally AdS$_3\times {\rm S}^3\times\mathbb{T}^4$ -- clearly then we have reproduced the D3-D3 near-horizon\footnote{I.e. the D1-D5 near-horizon twice T-dualised on different U(1)'s within $\mathbb{T}^4$.}, one can check that the flux is consistent with this. There are no further solutions with $\kappa=0$.

\subsubsection{S$^2$ case}\label{sec:S2}
In this section we shall study the case $\kappa > 0$, i.e.\ the two-sphere case. When $\kappa\neq 0$, positivity of the Riemann surface factor in \eqref{eq:metsol} demands that
\be
-\frac{\kappa}{1-\epsilon^2}>0\, ,
\ee
so when $\kappa>0$ we must have $\epsilon^2>1$, the symmetry of the solution means we can take this to be $\epsilon>1$ and so $m^-<0$ without loss of generality. Turning our attention to \eqref{eq:emedding} we see that when the right-hand side is positive/negative  it becomes an embedding equation for dS$_2$/AdS$_2$, both of which are non-compact and pseudo-Riemannian -- however $(\mu_{\pm},\,\mu_0)$ do not appear in the definition of ${\cal M}_5$ with the correct signs to give rise to sub-manifolds of these topologies and in fact ${\cal M}_5$ remains positive in both these cases. One can parametrise
\begin{align}
&c^+ b^- m^-+c^- b^+ m^+=+1\,:~~~~c^+c^-\mu_0^2 =\cos^2\beta \cosh^2 r\,,~~~~  m^+ \mu^2_+=\sin^2\beta \cosh^2 r\, ,\nn\\[2mm]
&c^+ b^- m^-+c^- b^+ m^+=-1\,:~~~~c^+c^-\mu_0^2 =\cos^2\beta \sinh^2 r\, ,~~~~  m^+ \mu^2_+=\sin^2\beta \cosh^2 r\, ,
\end{align}
without loss of generality, however in either case  $0\leq r<\infty$, and the upper limit is at infinite proper distance in the ten-dimensional space, so the solutions following from these tunings are unbounded. There is of course one final option, fix
\be
c^+ b^- m^-+c^- b^+ m^+=0\, .
\ee
Here one can parameterise
\be
c^+c^-\mu_0^2 =r^2\cos^2\beta,\quad  m^+ \mu^2_+=r^2 \sin^2\beta \, ,
\ee
but once again $0\leq r<\infty$, with the upper bound at infinite proper distance. Hence while solutions exist with S$^2$, they do not represent good holographic duals to CFTs.
\subsubsection{H$^2$ case}\label{sec:genMN}
In this section we shall study the case $\kappa < 0$,  where the Riemann surface is a compact quotient of H$^2$. Positivity of the metric in this case requires that $\epsilon^2<1$, meaning that $m^{\pm}>0$ and so \eqref{eq:emedding} embeds a two-sphere in $\mathbb{R}^3$. We thus fix
\be
c^+ b^- m^-+c^- b^+ m^+=2 c^+ c^-\,,~~~~~ m^+\mu_+^2=2 c^+ \cos^2\alpha\cos^2\beta\,,~~~~m^-\mu_-^2=2 c^- \cos^2\alpha\sin^2\beta\nn\,,
\ee
without loss of generality which makes the warp factor become
\be
\Lambda= 1+\sin^2\alpha+\epsilon \cos^2\alpha \cos(2\beta)\, ,
\ee
which has no zeros as $-1<\epsilon<1$. The internal five-manifold on the other hand can be written as
\begin{align}
\dd s^2({\cal M}_5) &= 2 \frac{\Lambda}{{\cal Q}}\dd\alpha^2+2  \sin^2\alpha \dd\psi_2^2+4\cos^2\alpha \bigg( \frac{{\cal Q}}{1-\epsilon^2} D\beta^2+\frac{1}{1+\epsilon} \cos^2\beta D\tau^2_++ \frac{1}{1-\epsilon} \sin^2\beta D\tau^2_-\bigg)\, ,\nn\\[2mm]
 {\cal Q}&\equiv1+\epsilon \cos(2\beta)\, ,~~~~D\tau_{\pm}\equiv\dd\tau_{\pm}+\frac{V}{2}\,,~~~~D\beta\equiv \dd\beta-\epsilon\frac{\tan\alpha \sin(2\beta)}{\cal Q}\dd\alpha\, ,
\end{align}
which is topologically a five-sphere with $0\leq\alpha<\frac{\pi}{2},~0\leq\beta<\frac{\pi}{2},~0\leq\psi_2<2\pi$ and the U(1) directions $\tau_{\pm}$ are defined as
\be
2\dd\tau_{\pm}\equiv(1\pm \epsilon)\dd\psi_1\pm \dd\phi\, ,
\ee
and have period $2\pi$. Notice that when $\epsilon=0$ we have $D\beta=\dd\beta$ and the metric becomes that of the AdS$_3$ solution of \cite{Maldacena:2000mw}. The general solution is a parametric deformation breaking the U(1)$\times$U(1)$\times$SU(2) isometry to U(1)$^3$, whilst preserving ${\cal N}=(2,2)$ supersymmetry. One can show that the two form flux is given by
\begin{align}
F^{(2)}&=-2 \sin(2\alpha) \dd\alpha\wedge\bigg((1-\epsilon)\frac{\cos^2\beta}{{\cal Q}}\dd\tau_++(1+\epsilon)\frac{\sin^2\beta}{{\cal Q}} \dd\tau_-+\left(\frac{1}{{\cal Q}}-\frac{1}{2}\right)V\bigg)\\[2mm]
&-2 \cos^2\alpha \sin(2\beta)D\beta\wedge\bigg(\frac{1-\epsilon}{1+\epsilon}\dd\tau_+-\frac{1+\epsilon}{1-\epsilon}\dd\tau_-- \frac{2\epsilon}{1-\epsilon^2}V\bigg)\nonumber\\
&+ e^{2A_0}\bigg(\frac{2\Lambda}{(1-\epsilon^2)}+\cos^2\alpha \bigg)\dd X_1\wedge \dd X_2\, .\nn
\end{align}

\subsubsection{Flux quantisation}
In order to compare to CFT quantities we must first quantise the flux. To this end it is helpful to know the internal part of the five-form flux explicitly,  we find
\begin{align}
\frac{1}{L^4}F^{\text{int.}}_5&=\dd\psi_2\wedge\bigg[8\frac{\sin\alpha \cos^3\alpha}{1-\epsilon^2}\frac{\sin(2\beta)}{\Lambda^2}\bigg(2\Lambda+ (1-\epsilon^2)\cos^2\alpha\bigg)\dd\alpha\wedge D\beta \wedge D\tau_-\wedge D\tau_+\nn\\[2mm]
&+2\sin(2\alpha)e^{2A_0}\dd X_1\wedge \dd X_2\wedge \dd\alpha\wedge \bigg(\frac{\cos^2\beta}{(1-\epsilon){\cal Q}}\dd\tau_++\frac{\sin^2\beta}{(1+\epsilon){\cal Q}}\dd\tau_-\bigg)\nn\\[2mm]
&+\frac{\sin^2(2\alpha)\sin(2\beta)}{(1-\epsilon^2)\Lambda}e^{2A_0}\dd X_1\wedge \dd X_2\wedge D\beta\wedge (\dd\tau_+-\dd\tau_-)\bigg]\, .
\end{align}
The relevant part of this is the first line. We should impose
\be\label{Fluxquant}
\frac{1}{(2\pi)^4}\int_{{\cal M}_5}F^{\text{int.}}_5= N \,,
\ee
in units where $g_s=\alpha'=1$. The integral is non-trivial but can be performed exactly,  ultimately we find that we must tune
\be
\frac{2L^4}{\pi(1-\epsilon^2)}= N \, .
\ee

\subsubsection{Central charge }\label{sec:central}

In this section we will study the central charge with which we may compare to the field theory results presented in section \ref{sec:SCFT}. Since the R-charges may be obtained from a specification of the more general solution we present later in section \ref{sec:gen-observables}, we will suppress the calculation of them here. The Brown--Henneaux formula \cite{Brown:1986tm}
\be
c_{\text{sugra}}= \frac{3}{2 G_{N}^{(3)}}
\ee
computes the sum of the left- and right- moving central charges of the field theory
\be
c_{\text{sugra}}=\frac{c_{L}+c_{R}}{2}\, .
\ee
Since the theories we are considering preserve $\mathcal{N}=(2,2)$ supersymmetry we have $c_{L}=c_{R}$, and therefore the gravitational anomaly should vanish. The dimensionally reduced Newton's constant appearing in the Brown--Henneaux formula is computed by reducing the full 10d solution on the compact internal manifold down to 3d. 
For a type II AdS$_3$ solution with metric of the form
\be
\dd s^2= e^{2A_3}\dd s^2(\text{AdS}_3)+ \dd s^2(\mathcal{M}_7)\, ,
\ee
the central charge is given by\footnote{This expression is in Einstein frame, though we have set the dilaton to zero, so this is identical to the string frame for us.}
\be\label{eq:centralcharge}
c_{\text{\text{sugra}}}= \frac{3}{2^4  \pi^6}\int_{\mathcal{M}_7}e^{A_3}\dd\text{vol}(\mathcal{M}_7)\, .
\ee 
For the H$^2$ solution we have
\be
e^{A_3}= L \Lambda^{\frac{1}{4}}\, ,\quad \dd\text{vol}(\mathcal{M}_7)= \frac{16 L^7}{1-\epsilon^2} \Lambda^{-\frac{1}{4}}e^{2A_0}\dd X_1\wedge \dd X_2\wedge\dd \text{vol}(\text{S}^5)\, ,
\ee
where by $\dd\text{vol}(\text{S}^5)$, we mean the volume form on the unit norm round five-sphere, integrating to $\pi^3$.  We thus find
\be
c_{\text{sugra}}= 3 N^2(\mathfrak{g}-1)\, ,
\ee
where we have used that the volume of the Riemann surface is given by
\be
\vol(\Sigma_{\mathfrak{g}})=\int_{\Sigma_\mathfrak{g}} \me^{2 A_0}\dd X_1\wedge \dd X_2=\int_{\Sigma_{\mathfrak{g}}} \rho=4 \pi (\mathfrak{g}-1)\, ,
\ee
where $\rho$ is the Ricci form, which follows since the chosen metric on the Riemann surface is Einstein.

\subsection{A more general solution}\label{sec:gen}
In this section we will generalise the previous solution by adding two additional free parameters.
We generalise the previous solution by taking as ansatz for $I$ 
\begin{align}
I=-\mu_0^2 (\log\mu_0^2-1)-\frac{1}{2R^+}\mu_+^2(\log \mu_+^2-1)-\frac{1}{2R^-}\mu_-^2 (\log \mu_-^2-1)+2 \nu(t^++t^-)\, ,
\end{align}
where
\be
\mu_{\pm}^2=b^{\pm}-R^{\pm}\left[(1+r^\pm)t^{+}+(1-r^{\pm}) t^{-}\right]\, .
\ee
Note that the previous ansatz for $I$ is recovered for $r^{+}=-r^{-}=1$ and $c^{\pm}=2R^{\pm}$.
As before the solution is split into distinct cases depending on the genus. The universal sector fixes the constants
\begin{align}
a^{+}= \frac{\kappa}{4}(2+r^{+}+r^{-})\, , \quad m^{\pm}=\frac{1\pm \epsilon}{2}\, ,\quad R^{\pm}\neq 0
\end{align}
whilst the curvature dependent part satisfies
\begin{align}\label{eq:topologydepcondsgen}
\kappa&=0: \quad \epsilon=r^{\mp}\, ,\quad c= -8 b^{\mp} R^{\pm}(r^{\pm}-\epsilon)^2 \me^{4 \nu}\, ,\\
\kappa&\neq 0:\quad \me^{4 \nu}=\frac{\kappa}{16 R^{+}R^{-}(\epsilon-r^{-})(\epsilon-r^{+})}\, ,\nonumber \\
&\quad\quad \epsilon=\frac{c R^{+}R^{-}(r^++r^-)+\kappa(b^{+}R^{-}r^{-}+b^{-}R^{+}r^{+})\pm \sqrt{R^+ R^-(r^+-r^-)^2(c^2 R^+ R^--b^-b^+)}}{\kappa(b^{+}R^-+b^- R^{+})+2 c R^+ R^-}\nonumber\, .
\end{align}
Note that these reduce to \eqref{eq:topologydepconds} for $r^{+}=-r^{-}=1$ and $c^{\pm}=2R^{\pm}$ as they should. Moreover, we have a similar constraint on the functions $\mu_{0}$, $\mu_{\pm}$ as in \eqref{eq:emedding} given by
\be\label{eq:emeddinggen}
2R^+ R^- (r^+-r^-)\mu_0^2+R^- (\epsilon-r^-)\mu_+^2+R^-(r^+-\epsilon)\mu_-^2=b^- R^+(r^+-\epsilon)+b^+ R^-(\epsilon-r^-)\, .
\ee
As before, this defines an embedding equation of a surface into a three-dimensional ambient space. Note that the inclusion of the parameters $r^{\pm}$ shift the coefficients of the functions $\mu_{\pm}$ so that it is no longer simply $m^{\pm}$ as in \eqref{eq:emedding}. We may now assemble the full metric of the solution. We have
\begin{align}
\frac{1}{L^{2}}\dd s^2 &=\sqrt{\Lambda} \Big[ \dd s^2 (\text{AdS}_3)+4 R^{+}R^{-}(r^+-r^-)^2\me^{4\nu} \me^{2 A_0}(\dd X_1^2+\dd X_2^2)\Big]\label{eq:metgen}\\
&+\frac{1}{\sqrt{\Lambda}}\Big[\dd \mu_0^2+\frac{1}{2 R^+}\dd \mu_+^2+\frac{1}{2 R^-}\dd \mu_-^2+\Lambda \eta^2 +\frac{ \mu_-^2\mu_+^2}{2(R^-(r^--\epsilon)^2 \mu_+^2+R^+(r^+-\epsilon)^2 \mu_-^2)}D\phi^2\nonumber\\
&+\frac{\mu_0^2 [R^-(r^--\epsilon)^2 \mu_+^2 +R^+ (r^+-\epsilon)^2 \mu_-^2]}{2R^+R^- (r^+-r^-)^2 \mu_0^2+R^- (r^--\epsilon)^2 \mu_+^2+R^+(r^+-\epsilon)^2 \mu_-^2}D\phi_2^2\Big]\nonumber\, ,
\end{align}
where
\begin{align}
D\phi&\equiv\dd \phi-\left(\epsilon-\frac{r^++r^-}{2}\right) \kappa V\, ,\nonumber\\
D\phi_2&\equiv \dd \phi_2+\kappa V - \frac{R^-(r^--\epsilon)\mu_+^2+R^+(r^+-\epsilon)\mu_-^2}{R^-(r^--\epsilon)^2\mu_+^2+R^+ (r^+-\epsilon)^2 \mu_-^2}D\phi\, ,\\
\eta&\equiv\frac{1}{2} \left(\dd \phi_1+\frac{ R^+(r^+-\epsilon)^2 \mu_-^2+R^-(r^--\epsilon)^2 \mu_+^2}{ R^+ R^-(r^+-r^-)^2 \Lambda}D\phi_2\right)\, ,\nonumber
\end{align}
and
\be
\Lambda\equiv\mu_0^2+\frac{(r^--\epsilon)^2 }{2 R^+(r^+-r^-)^2}\mu_+^2+\frac{(r^+-\epsilon)^2}{2 R^-(r^+-r^-)^2}\mu_-^2\, .
\ee
From the presentation above we see that the metric looks locally like a U$(1)^3$ preserving deformed five-sphere. In addition, from the $\mathcal{N}=(0,2)$ perspective of section \ref{sec:(0,2)}, one can identify the one-form $\eta$ as the one-form dual to the R-symmetry vector. This puts the metric on the five-sphere in the form of a U$(1)$-fibration over $\mathbb{CP}^2$ equipped with a non-Einstein metric. The two-form flux is given by 
\begin{align}
F^{(2)}=2 J_6-\dd (\Lambda \eta)\, ,
\end{align}
with
\begin{align}
J_6&=4 \me^{4 \nu}R^+R^- (r^+-r^-)^2 \Lambda \dd V-\frac{1}{4}\left[(1-\epsilon)\dd t^-+(1+\epsilon)\dd t^+\right]\wedge D\phi_{2}\nonumber\\
&+\frac{(1+r^+)R^+(r^+-\epsilon)\mu_-^2-(1+r^-)R^-(\epsilon-r^-)\mu_+^2}{4(R^+(r^+-\epsilon)^2 \mu_{-}^2 +R^-(\epsilon-r^-)^2 \mu^2_{+})}\dd t^+\wedge D\phi\nonumber\\
&+\frac{(1-r^+)R^+(r^+-\epsilon)\mu_-^2+(r^--1)R^- (\epsilon-r^-) \mu_+^2}{4(R^+(r^+-\epsilon)^2 \mu_{-}^2 +R^-(\epsilon-r^-)^2 \mu^2_{+})}\dd t^-\wedge D\phi\, .
\end{align}

\subsubsection{$\mathbb{T}^2$ case}

As in section \ref{sec:T2} the $\mathbb{T}^2$ example reduces to the AdS$_3\times {\rm S}^3 \times \mathbb{T}^4$ solution, as such we shall suppress further analysis of this case.

\subsubsection{S$^2$ case}

Let us turn to the S$^2$ case. It is clear that we require $R^{\pm}>0$ and therefore positivity of the warp factor of the Riemann surface, $\me^{4 \nu}$ implies that
\be
(\epsilon-r^-)(\epsilon-r^+)>0\, .
\ee
Studying the embedding equation \eqref{eq:emeddinggen} we see that it defines an embedding into either three-dimensional anti-de Sitter, Minkowski or de Sitter space for choices of the parameters. In each case the solutions will be non-compact and therefore they do not give rise to good holographic duals as in section \ref{sec:S2}.

\subsubsection{H$^2$ case}
Instead, let us consider the $\mathfrak{g}>1$ case. Positivity of the warp factor requires 
\be
(r^+-\epsilon)(r^--\epsilon)<0\, .
\ee
Due to the symmetry of the solution we may assume without loss of generality that $r^+>r^-$; if this is not the case then we may relabel as $+\leftrightarrow -$. We see that we must therefore require 
\be
r^+>\epsilon\, ,\quad r^-<\epsilon\, ,
\ee
and consequently the embedding equation is that of a two-sphere in $\mathbb{R}^3$. We may then parametrise as
\begin{align}
\mu_{0}^2&=\frac{b^{-}R^{+}(r^{+}-\epsilon)+b^{+}R^{-}(\epsilon-r^-)}{2 R^+R^-(r^+-r^-)} \sin^2 v\, , \quad \mu_-^2= \frac{b^- R^+(r^+-\epsilon)+R^- b^+ (\epsilon-r^-)}{R^+ (r^+-\epsilon)} \cos^2 v\cos^2 u\, ,\nonumber\\
 \mu_{+}^2&=\frac{b^{-}R^{+}(r^{+}-\epsilon)+R^- b^{+}(\epsilon-r^-)}{R^- (\epsilon-r^-)} \cos^2 v \sin^2 u\, ,
\end{align}
giving rise to a compact internal manifold.

\subsubsection{Toric data and three-cycles}

We now want to analyse the regularity of these generalised solutions. We saw earlier that regularity of the metric can be ensured by correctly identifying the embedding coordinates. In this section we will take an alternative approach using the toric data of the five-sphere at fixed point on the Riemann surface. Note that the compact part of the metric in \eqref{eq:metgen}, without the Riemann surface, has been written as a U$(1)^3$ fibration over a two-dimensional space defined by the embedding condition \eqref{eq:emeddinggen}. One can then use the tools of toric geometry to analyse the regularity of the metric. Moreover, this analysis allows us to identify the three-cycles, each corresponding to a degeneration surface of the metric, on which one may wrap D3-branes giving rise to di-baryonic operators of the dual SCFT. Of course, since on the five-sphere $b_3=0$ defining such charges is subtle. From the toric viewpoint these three-cycles wrapped by D3-branes are obtained when one goes to the edge of the polytope over which the U$(1)^3$ fibration is defined and where one of the U$(1)$'s shrinks. 
 With this in hand we may compute the R-charges of these operators and compare to the field theory. 
In the following we will consider just the $\kappa\neq0$ case since the $\kappa=0$ case reduces to the previously studied AdS$_3\times \text{S}^3\times \mathbb{T}^4$ solution. 

The form of the ansatz for the function $I$ is reminiscent of the canonical potential for a symplectic toric manifold. Here, the $\mu_\bullet^2$'s play the role of the functions defining the edges of the polytope. Note that the metric has a singular like behaviour when any of $\mu_0$ or $\mu_{\pm}$ vanish, the goal is to obtain constraints such that the degeneration is smooth.\footnote{For a more detailed discussion as to why the following works see \cite{Couzens:2019mkh} for example. }
To this end we construct Killing vectors which have zero norm on some locus of the solution. The degeneration locus is associated to the edges of a 2d polytope over which the U$(1)^3$ is fibered. Requiring that the surface gravity of these Killing vectors is normalised to 1 on the respective degeneration surface, and giving the dual coordinate period $2\pi$ leads to a smooth degeneration and a regular metric. In the case at hand there are three Killing vectors, which, after the prescribed normalisation condition discussed above, are given by
\be
k_{0}=2 \partial_{\phi_{1}}-\partial_{\phi_2}\, ,\quad k_{\pm}=(r^\pm-\epsilon)\partial_{\phi}+\partial_{\phi_2}\, .
\ee
Defining a new $2\pi$-periodic coordinate for each of these three vectors via, $k_{\bullet}=\partial_{\psi_{\bullet}}$, we find that the metric is regular. In order to compute various integrals it is useful to perform a change of coordinates,
\begin{align}
\phi_{1}= 2 \psi_{0}\, ,\quad \phi_{2}=-\psi_0+\psi_++\psi_-\, ,\quad  \phi=(r^+-\epsilon)\psi_++(r^--\epsilon) \psi_-\, .
\end{align}
These new $2\pi$-periodic coordinates give a free action for the U$(1)^3$ torus action.

To extend this regularity analysis to the full manifold we now need to make sure that the fibration over the Riemann surface is well defined. Viewing the fibration as a gauging of the U(1)$^3$ torus action via
 \be
 \dd \psi_\bullet\rightarrow D\psi_\bullet \equiv \dd \psi_\bullet+A_\bullet\, ,
 \ee
 with the $A_\bullet$ gauge fields on the Riemann surface. The fibration is well-defined provided that the field-strength of the gauge fields $A_\bullet$ over the Riemann surface have integral period. Explicit computation gives
 \be
 A_0=0\, ,\quad A_+=A_-= \frac{\kappa}{2}\, V\, ,
 \ee
and therefore the fibration is well-defined since
\be
\frac{1}{2\pi}\int_{\Sigma_{\mathfrak{g}}}\dd V=2(\mathfrak{g}-1)\, .
\ee
From the non-trivial gauge fields we can read off the topological twist of the dual field theory. We see that the topological twist is performed by turning on a background gauge field of the form
\be
A=\frac{\kappa}{2} (T_++T_-)V\, . 
\ee
Finally we may identify the (right-moving) R-symmetry of the solution from the expression for the one-form $\eta$. We find
\be
{\rm R}_{R}=\partial_{\psi_0}+ \frac{r^--\epsilon}{r^+-r^-}\partial_{\psi_+}+\frac{\epsilon-r^+}{r^+-r^-}\partial_{\psi_-}\, .\label{Rsymvec}
\ee
Note in particular that the R-symmetry does \emph{not} mix with the isometries of the Riemann surface. This is a distinction between the solutions discussed in this section and the ones in the following section \ref{sec:disk}.

\subsubsection{Flux quantisation and observables}\label{sec:gen-observables}

Having discussed the geometry of the solution and the various three-cycles let us first consider the quantisation of the five-form flux. We must ensure that this is an integer over all integral five-cycles in the geometry according to the quantisation condition \eqref{Fluxquant}. Since the geometry is a five-sphere fibered over a Riemann surface the only integral five-cycle is the five-cycle at fixed point on the Riemann surface. The internal part of the five-form is
\begin{align}
\frac{1}{L^{4}}F_5^{\text{int.}}&=\frac{(r^+-r^-)(b^+ R^-(\epsilon-r^-)+b^- R^+ (r^+-\epsilon))^2}{2 (R^+ R^-)^2(r^+-r^-)(r^+-\epsilon)(\epsilon-r^-)\Lambda^2}\cos u \sin u \cos^3 v\sin v \times\nonumber\\
&\Big(\Lambda+\cos^2 v \frac{(r^+-\epsilon)(\epsilon-r^-)\left[b^+ R^-(\epsilon-r^-)+b^- R^+(r^+-\epsilon)\right]}{2 R^+ R^- (r^+-r^-)^3}\Big)\dd u \wedge \dd v \wedge \dd\psi_0\wedge \dd \psi_+\wedge \dd \psi_-\nonumber\\
&+\dots\, ,
\end{align}
where the suppressed terms do not have support on the topological five-sphere.
Integrating the flux over the five-sphere we find the quantisation condition
\be
N=\frac{L^4}{8 \pi}\frac{(r^+-r^-)[b^+ R^-(r^--\epsilon)-b^-R^+(r^+-\epsilon)]}{ R^+ R^-(r^+-\epsilon)(\epsilon-r^-)}\, .
\ee

We can now compute the R-charges of baryonic operators in the SCFT. These correspond to wrapping D3-branes on three-cycles on the five-sphere. The three-cycles, let us denote them by $S_\bullet$, are the ones obtained by going to the facet of the polytope, upon which a U$(1)$ shrinks.\footnote{We emphasise again that these three-cycles are not elements of $H^3($S$^5,\mathbb{Z})$ since $b_3($S$^5)=0$.} Following \cite{Couzens:2017nnr} (see also \cite{Couzens:2018wnk}) the R-charge is given by 
\be
{\rm R}[S_\bullet]=\frac{L^4}{(2\pi)^3 }\int_{S_\bullet}\me^{4\Delta}\dd \vol(S_\bullet)=\frac{L^4}{(2\pi)^3 }\int_{S_\bullet}\eta\wedge J\, ,
\ee
where the last inequality has been written using $(0,2)$ language. Explicit computation gives 
\begin{align}
{\rm R}[S_0]&=-\frac{L^4}{8\pi} \frac{(r^+-r^-)[b^+R^-(r^{-}-\epsilon)-b^- R^+(r^+-\epsilon)]}{R^+ R^- (r^+-\epsilon)(r^{-}-\epsilon)}\, ,\\
{\rm R}[S_+]&=\frac{L^4}{8\pi } \frac{b^+ R^-(r^--\epsilon)-b^- R^+ (r^+-\epsilon)}{R^+ R^- (r^+-\epsilon)}\, ,\nonumber\\
{\rm R}[S_-]&=\frac{L^4}{8\pi }\frac{b^- R^+(r^+-\epsilon)-b^+ R^- (r^{-}-\epsilon)}{R^+ R^- (r^--\epsilon)}\, ,\nonumber
\end{align}
which after using the definition of $N$ become
\be
{\rm R}[S_0]=N\,,\quad {\rm R}[S_+]=\frac{N(\epsilon-r^-)}{r^+-r^-}\, ,\quad {\rm R}[S_-]=\frac{N(r^+-\epsilon)}{r^+-r^-}\, .
\ee
Note that the R-charges are all positive and moreover they sum to $2 N$ as expected \cite{Couzens:2018wnk},
\be
{\rm R}[S_0]+{\rm R}[S_+]+{\rm R}[S_-]=2N\, .
\ee
In addition, from the form of the R-symmetry vector as given in \eqref{Rsymvec}, it is clear that the dual operators in the field theory have charge 1 under the associated degenerating Killing vector, and vanishing charge under the others directions. We may compare this with \eqref{FTRcharges} and find agreement upon identifying 
\be
\cos^2 \alpha_{R}=\frac{\epsilon-r^-}{r^+-r^-}\, .
\ee

Finally, let us compute the central charge. As before we use the Brown--Henneaux formula as given in \eqref{eq:centralcharge}, and as expected from the field theory analysis, the result is
\begin{align}
c=3 (\mathfrak{g}-1) N^2\, .
\end{align}
Note in particular that the newly introduced parameters drop out of the central charge, despite not dropping out of the R-charges of the fields.

\section{D3-branes wrapping a topological disk}\label{sec:disk}
In \cite{Boido:2021szx} a solution with D3-branes wrapping a spindle was obtained in 5d U(1)$^3$ gauged supergravity and then lifted to ten dimensions. The result locally coincides with a solution originally derived in \cite{Gauntlett:2006ns}, but the interpretation in terms of a spindle and CFT analysis is new. The solution supports multiple charges and generically preserves ${\cal N}=(0,2)$ supersymmetry, however, as we shall establish in this section, for a certain tuning of these charges it experiences an enhancement to ${\cal N}=(2,2)$, where the D3-branes now actually wrap a Riemann surface with the topology of a disk and includes additional source D3-branes. This solution can be embedded in the canonical form of section \ref{sec:generalform}, interestingly the Riemann surface which the branes wrap is not that of the $(X_1,\,X_2)$ directions. Instead, as we establish in section \ref{sec:neq22proof}, $(X_1,\,X_2)$ span a two-sphere in the co-dimensions of the wrapped D3-branes.  \\

\subsection{Summary of \cite{Boido:2021szx}}
The metric of the solution of \cite{Boido:2021szx} takes the form\footnote{The flux can be found in this reference.}
\begin{align}
\frac{1}{L^2}\dd s^2&= \sqrt{W}H(x)^{\frac{1}{3}}\left[\dd s^2(\text{AdS}_3)+ \dd s^2(\Sigma_2)\right]+ \frac{1}{\sqrt{W}}\sum_{I=1}^3(X^I)^{-1}\left[\dd\mu^2_I+\mu^2_I(\dd\phi_I+A_I)^2\right]\,,\nn\\[2mm]
W&=\sum_{I=1}^3 X^I\mu_I^2\,,~~~~A_I= \frac{x-x_0}{x+3K_I}\dd\varphi\,,~~~~ X^I=\frac{H(x)^\frac{1}{3}}{x+3K_I}\,,\nn\\[2mm]
\dd s^2(\Sigma_2)&= \frac{1}{4 P(x)}\dd x^2+ \frac{P(x)}{H(x)}\dd\varphi^2\,,
\end{align}
where $\mu_I$ embed a unit radius two-sphere into $\mathbb{R}^3$ and the functions of $x$ are 
\be
H=(x+3K_1)(x+3K_2)(x+3K_3)\,,\quad P= H-(x-x_0)^2\,,
\ee
where \cite{Boido:2021szx}  constrain $K_1+K_2+K_3=0$, which kills the $O(x^2)$ term in $H$. The $x$ direction can then be bounded between two real roots of the cubic polynomial $P$ for which $H\neq 0$, and the surface becomes $\Sigma_2= \mathbb{WCP}^1_{[n_+,n_-]}$, which is an orbifold known as a spindle. Here $n_{\pm}\in \mathbb{N}$, are related to the period of $\varphi$. When $K_1=K_2=K_3=0$  the solution of \cite{Ferrero:2020laf} is recovered. In this section we shall consider a different tuning of these parameters, not considered in \cite{Boido:2021szx}.

\subsection{An ${\cal N}=(2,2)$ tuning}
We would like to extract an ${\cal N}=(2,2)$ solution from the local solution of the previous section. The first thing to appreciate is that $K_1+K_2+K_3=0$ is not a requirement for supersymmetry (at least in ten dimensions), so let us instead tune the parameters as
\be\label{eq:neq11tuning}
K_2=K_1=K\, , \quad K_3=-\frac{1}{3}x_0\, .
\ee
The connection $A_3$ then becomes topologically trivial, $\{A_1=A_2,\, X_1=X_2=X_3^{-2}\}$ and so the U(1)$^3$ isometry of the five-dimensional internal space gets enhanced to SU(2)$\times$U(1) -- similar to the ${\cal N}=(2,2)$ solution of \cite{Maldacena:2000mw}. This solution experiences an enhancement of supersymmetry to ${\cal N}=(2,2)$, as we prove in the following section, \ref{sec:neq22proof}. The two functions defining the solution become
\be
H=(3K+x)^2(x-x_0)\, , \quad P= (x-x_0)Q\, , \quad Q=(3K+x)^2-x+x_0\,.
\ee
For certain tunings of $(K,x_0)$ the polynomial $Q$ contains two real roots $x=x_{\pm}$ namely 
\be\label{eq:xpm}
x_{\pm}= \frac{1}{2}\left(1-6K\pm \sqrt{1-12K-4x_0}\right)\,.
\ee
$\Sigma_2$  exhibits spindle-like behaviour between these loci, however a physical metric should be real and positive which requires $\{Q>0,~x>x_0,~3 K+ x >0\}$, and $Q$ is strictly negative for $x_-< x<x_+$. It is however possible to achieve a physical metric and to bound $x$ to the interval $[x_0,x_-]$ provided that $-\frac{1}{3}x_0<K<\frac{1}{12}(1-4 x_0)$. At $x=x_0$  $H$ and $P$ both vanish, but $P/H$ remains finite, so similar to the solution of \cite{Bah:2021hei}, the topology of $\Sigma_2$ is now that of a disk with $\mathbb{R}^2/\mathbb{Z}_k$ orbifold singularity at $x=x_-$.  Close to $x=x_-$ only $P$ exhibits a zero, defining $u= (x_--x)$ we find
\be
\dd s^2(\Sigma_2)= \frac{1}{x_+-x_-}\bigg(\dd u^2+ \frac{(x_+-x_-)^2(x_--x_0)^2}{H(x_-)} u^2 \dd\varphi^2\bigg)\, ,
\ee
at this loci, so that we have a  $\mathbb{R}^2/\mathbb{Z}_k$ orbifold when the period of $\varphi$ is $\Delta\varphi= \frac{2\pi \sqrt{H(x_-)}}{k(x_+-x_-)(x_--x_0)}$ for $k\in\mathbb{N}$ or a regular zero for $k=1$.  The behaviour close to $x=x_0$ in the full space needs more care to disentangle: first one should note that warp factor $W$ can now be written as
\be
H^\frac{2}{3}W= (3K+x)\bigg((x-x_0)+ (3K+x_0)\mu_3^2\bigg)\, ,
\ee
which provided $\mu_3\neq 0$, remains finite as $x$ approaches $x_0$. Similarly around $x=x_0$ when $\mu_3\neq 0$ we find $W (X^{1,2})^{-2}= \mu_3^{-2}$ and $W(X^3)^{-2}=(x-x_0)^2(3K+x_0)^{-2}\mu_3^{-2}$. We see then that the sub-metric spanned by $(x,\,\phi_3+\varphi)$ vanishes as $\mathbb{R}^2$ in polar coordinates. The behaviour as both $\mu_3,(x-x_0)\to 0$ is a bit subtle, one can study it by changing coordinates to
\be
\mu_3^2= r\cos^2\left(\frac{\tilde\theta}{2}\right),~~~~\mu_2+\text{i} \mu_1= e^{\frac{\text{i}}{2} \beta}\sqrt{1-\mu_3^2}\,,~~~~ x=x_0+ (3K+x_0)r \sin^2\left(\frac{\tilde\theta}{2}\right)\, ,
\ee
and then expanding about $r=0$. To leading order the metric then becomes
\be
\frac{1}{L^2}\dd s^2= \sqrt{r}(3K+x_0)\left[\dd s^2(\text{AdS}_3)+\dd\varphi^2\right]+\frac{1}{4 \sqrt{r}}\left[\dd r^2+r^2 \dd s^2(\tilde{\text{S}}^2)+ 4\dd s^2(\text{S}^3)\right]\, ,
\ee
where $(\tilde\theta,\,\phi_3+\varphi)$ span the two-sphere and $(\beta\,,\phi_1,\,\phi_2)$ span the three-sphere (in topological joint coordinates). It is not hard to see that this behaviour is singular, it is in fact that of a partially localised stack of D3-branes  with worldvolume $(\text{AdS}_3,\,\varphi)$ that  are smeared over S$^3$. Thus we see that at generic points of the deformed two-sphere spanned by $\mu_I$ the solution is bounded between a regular zero at $x=x_0$ and a $\mathbb{R}^2/\mathbb{Z}_k$ orbifold at $x=x_-$, with  flavour D3-branes at the loci $(x=x_0,\,\mu_3=0)$.

To further study the solution let us fix
\begin{align}
\mu_3 &=\sin\alpha\,,~~~~\mu_2+\text{i}\mu_1 =\cos\alpha e^{\frac{\text{i}}{2}\beta}\, ,\nn\\[2mm]
\phi_1&=-\frac{1}{2}(\tau_1-\tau_2)\,,~~~~\phi_2=\frac{1}{2}(\tau_1+\tau_2)\,,~~~~\phi_3=\psi- \varphi\, ,\nn\\[2mm]
\Lambda&=x-x_0+(3K+x_0)\sin^2\alpha\, ,
\end{align}
which allows us to write the metric as
\begin{align}
\frac{1}{L^2}\dd s^2&=\sqrt{3K+x}\sqrt{\Lambda}\left[ \dd s^2(\text{AdS}_3)+ \frac{1}{4(x-x_0)Q}\dd x^2+\frac{Q}{(3K+x)^2}\dd\varphi^2\right]+\frac{1}{\sqrt{\Lambda}} \dd s^2({\cal M}_5)\,,\nn\\[2mm]
\dd s^2({\cal M}_5)&=\frac{\Lambda}{\sqrt{3K+x}}\dd\alpha^2+\frac{x-x_0}{\sqrt{3K+x}}\sin^2\alpha \dd\psi^2\nn\\[2mm]
&+\frac{1}{4}\sqrt{3K+x}\cos^2\alpha\left[\dd\beta^2+\sin^2\beta \dd\tau_2^2+ \left(\dd\tau_1+\cos\beta \dd\tau_2+2 \frac{x-x_0}{3K+x}\dd\varphi\right)^2\right]\,,\label{eq:disksol}
\end{align}
where ${\cal M}_5$ is topologically a five-sphere when the coordinates are taken to have periodicity
\be
0\leq\alpha<\frac{\pi}{2},~~~~0\leq\beta<\pi,~~~~0\leq\psi<2\pi,~ ~~~0\leq\tau_1< 2\pi,~~~~ 0\leq\tau_2 <4\pi\, .
\ee

\subsection{Embedding the topological disk solution into section \ref{sec:generalform}}\label{sec:neq22proof}

Above we have constructed a solution by taking a limit of the multi-charge solution of \cite{Gauntlett:2006ns} and then analysed its regularity. In this section we prove that the solution does indeed preserve ${\cal N}=(2,2)$ supersymmetry as claimed. We shall do so by embedding it into the general form of section \ref{sec:generalform}. Much of this can be reverse-engineered by comparing \eqref{eq:disksol} to \eqref{eq:7manifold} and \eqref{eq:generalform}. This is sufficient to establish how the coordinates are related, extract the values of $(\partial^2_{\Theta}D,\,\partial_y D)$ and to identify the Riemann surface of the local ansatz, which surprisingly turns out not to be $\Sigma_2$. In fact it is not hard to show that the $(X_1,\,X_2)$ directions in \eqref{eq:generalform} correspond to the two-sphere spanned by $(\beta,\,\tau_1)$ in \eqref{eq:disksol}, as such this is another example in the constant curvature ansatz, our first with a two-sphere. We must identify the coordinates of \eqref{eq:generalform}  as
\begin{align}
X_1&=\tan\left(\frac{\beta}{2}\right)\cos\tau_1\, ,~~~~X_2=\tan\left(\frac{\beta}{2}\right)\sin\tau_1\, ,\nn\\[2mm]
y&=(x-x_0)\sin^2\alpha\, ,~~~~\Theta=3K+2 y-x_0-\Lambda\, ,\nn\\[2mm]
\psi_1&=-\varphi-\tau_1-\tau_2\, ,~~~~\psi_2=\psi\, ,~~~~ \phi=\tau_1+\tau_2\, ,
\end{align}
and decompose the potential defining the system as
\be
D= I(y,\Theta)+(2y-\Theta)A_0\, ,~~~~ A_0= -\log\left(\frac{1}{2}\left(1+X_1^2+X_2^2\right)\right)\, ,
\ee
so that the partial differential equation of \eqref{MAeq} reduces to
\be\label{eq:IIII}
\partial_{y}^2I\partial_{\Theta}^2I-(\partial_y\partial_{\Theta}I)^2+\frac{(2y-\Theta)}{16 y^2}e^{-\partial_y I}=0\, .
\ee
It is possible to extract the following derivative of $I$ from the metric\footnote{We can also extract an expression for $\partial_y\partial_{\Theta}I$ which is consistent with these.}
\begin{align}
e^{\partial_y I}&=\frac{1}{32 \tilde{K}}\Delta_3\, ,\label{eq:partialyI}\\[2mm]
2\Delta_2\partial_{\Theta}^2I&=-\tilde{K}\Theta-(1-2\tilde{K})y+\frac{\Delta_2+\tilde{K}\sqrt{\Delta_1}}{2y-\Theta}-\frac{y(1-3\tilde{K}+y)}{\sqrt{\Delta_1}}-\frac{y(1-\tilde{K}-y)(\tilde{K}+y)}{(2y-\Theta)\sqrt{\Delta_1}}\,,\nn
\end{align}
where we introduce
\begin{align}
\tilde{K}&\equiv3K+x_0\, ,~~~~\Delta_1\equiv4\tilde{K}y+(\tilde{K}+y-\Theta)^2\, ,\nn\\[2mm]
\Delta_2&\equiv \tilde{K}^2+y(\Theta-1-y)+\tilde{K}[4y(1+y)-(1+4  y)\Theta+\Theta^2]\, ,\nn\\[2mm]
\Delta_3&\equiv2\Delta_2-(\tilde{K}+y)(\sqrt{\Delta_1}-2-y+\Theta)-\tilde{K}(2+\tilde{K}+5 y-2\Theta)\, .
\end{align}
The definitions \eqref{eq:partialyI} are self consistent (i.e.\ they give rise to the same $\partial_{\Theta}^2\partial_y I$) and are already enough to confirm that \eqref{eq:IIII} is satisfied, which is an important consistency check. Further, though non trivial, the  definition of $\partial_y I$ in \eqref{eq:partialyI} can be integrated to give $I$ up to a function of $\Theta$ which can then be fixed\footnote{Note that \eqref{eq:IIII} is invariant under $I\to I+f(\Theta)$ when $f''(\Theta)=0$, so we set terms linear in $\Theta$ to zero.} by imposing consistency with the expression for $\partial_{\Theta}^2I$. The final expression is rather complicated but of closed form, we find it convenient to decompose it as
\begin{align}
I&=I_0+I_++I_-+I_{\Theta}\, ,\nn\\[2mm]
I_{\Theta}&=\frac{1}{4}\left((\Theta-1)+\frac{1-2\tilde{K}-\Theta}{\sqrt{1-4 \tilde{K}}}\right)\log(\Theta-2\tilde{K}),~~~~I_{\pm}=\frac{1}{4\sqrt{1-4\tilde{K}}}q_{\pm}\left[\log\left(l_{\mp}+p_{\pm}\sqrt{\Delta_1}\right)-\log m_{\pm}\right]\, ,\nn\\[2mm]
I_0&=-y\big[1+\log(32\tilde{K}y)\big]+y\log\Delta_3-\Theta\bigg[\frac{1}{2}\log\bigg(\left(\tilde{K}+y+\sqrt{\Delta_1}\right)\left(\tilde{K}-y+\Theta+\sqrt{\Delta_1}\right)\bigg)-\log(2y-\Theta)\bigg]\nn\\[2mm]
&-\frac{1}{\sqrt{1-4\tilde{K}}}\left[2\tilde{K}-(1-\Theta)\left(1+\sqrt{1-4\tilde{K}}\right)\right]\log\Delta_2-\frac{1}{2}\log\left(3\tilde{K}+y-\Theta+\sqrt{\Delta_1}\right)\nn\\[2mm]
&-\frac{1-2\tilde{K}-\Theta}{2\sqrt{1-4\tilde{K}}}\log\left((1-\Theta)\left(1-\sqrt{1-4\tilde{K}}\right)-2y\sqrt{1-4\tilde{K}}-2\tilde{K}\right)\, ,\label{eq:complicatedIdef}
\end{align}
where $q_{\pm}\equiv q_1\pm q_2$ and so on for $p_\pm$, $l_\pm$ and $m_\pm$, with
\begin{align}
q_1&\equiv-1+2\tilde{K}+\Theta\,,~~~p_1\equiv1+\Theta-6\tilde{K}\,,~~~q_2\equiv\sqrt{1-4\tilde{K}}(1-\Theta)\,,~~~p_2\equiv\sqrt{1-4\tilde{K}}q_1\,,\nn\\[2mm]
l_1&\equiv\sqrt{1-4\tilde{K}}\big[\Delta_1+y(1+y)+\Theta(1+\tilde{K}+y)+\tilde{K}(\tilde{K}-2)\big]\,,~~~l_2\equiv(3\tilde{K}+y-\Theta)q_1\, ,\nn\\[2mm]
m_1&\equiv\sqrt{1-4\tilde{K}}(1+2y-\Theta)\,,~~~m_2\equiv q_1\, .
\end{align}
We have confirmed that \eqref{eq:complicatedIdef} does indeed solve \eqref{eq:IIII}, so the solution preserves ${\cal N}=(2,2)$ supersymmetry as claimed.

\subsection{Flux quantisation and central charge}

We now want to turn our attention to computing some observables of the dual CFT. 
We should first quantise the flux on ${\cal M}_5$, the relevant part is
\be
\frac{1}{L^4}F_5= \frac{1}{4}\sin\alpha\sin\beta \dd\alpha\wedge \dd\psi\wedge  \dd\beta \wedge \dd\tau_1\wedge \dd\tau_2+ \dots
\ee
where $\dots$ either vanishes on ${\cal M}_5$ or does so when it is integrated over it. We then have
\be
\frac{1}{(2\pi)^4}\int_{{\cal M}_5}F_5 = N~~~~ \Rightarrow~~~~ L^4=2\pi N\, .
\ee
We should also quantise the connections of the fibrations such that we have a well defined orbifold fibration, this demands that we impose that the field strengths $F_{I}=\dd A_{I}$ satisfy\footnote{Recall $dA_3=0$.},
\be\label{eq:qcond}
M/k=\frac{1}{2\pi}\int_{\Sigma_2}F_1=\frac{1}{2\pi}\int_{\Sigma_2}F_2= \frac{\Delta\varphi}{4\pi}[1-(x_+-x_-)]\,,~~~~ \Delta\varphi= \frac{\pi[1-(x_+-x_-)]}{k(x_+-x_-)\sqrt{x_--x_0}}\,,
\ee
for $M\in \mathbb{N}$, which fixes 
\be\label{eq:2branches}
3K+x_0= \frac{M(M\pm 1)}{(2M\pm 1)^2}\,.
\ee
Given the constraints on $(K,\,x_0)$ discussed below \eqref{eq:xpm} the positive branch is valid for $M\geq 1$ and the negative one for $M\geq 2$, so they are equivalent -- to be concrete we take the former.
We can compute the holographic central charge using \eqref{eq:centralcharge}, we find after substituting for $L$ and $3K+x_0$ this takes the form
\be
c_{\text{sugra}}= 3 N^2 \frac{M^2}{4k(1+2M)}\,, 
\ee
where again we expect the central charges of the dual CFT to be $c_L=c_R=c_{\text{sugra}}$. Clearly we have a deviation from the behaviour derived in section \ref{sec:SCFT}, but this should be no surprise as the D3-branes are now wrapping a topological disk which does not have constant curvature.

\subsection{Discussion}
In section \ref{sec:neq22proof} we showed how the solution is embedded into the canonical form of section \ref{sec:generalform}. To our knowledge this is the first time an example with branes wrapping a surface of non-constant curvature has been embedded into one of the various AdS classifications that are defined in terms of a Mong\'e--Ampere like equation -- it may be possible to construct similar solutions for other wrapped brane scenarios. Interestingly, for this example, the Riemann surface that the branes wrap and that of the classification are not the same -- indeed the $(X_1,\,X_2)$ directions of \eqref{eq:generalform} actually span a two-sphere. This is a novel feature of this solution and one should contrast this with the solutions in section \ref{sec:H2} where the Riemann surface of the classification and that wrapped by the D3-branes is identical.
This may provide a hint towards finding solutions with branes wrapping non-constant Riemann surfaces more broadly. It would be interesting to generalise this example as section \ref{sec:genMN} generalises \cite{Maldacena:2000mw}, but we shall not attempt that here. 

The dual field theory computation in this case is somewhat subtle. One cannot simply specialise the field theory result in \cite{Boido:2021szx} to our current case. Concretely, one finds that after $c$-extremization has been performed, despite the trial central charge having a well defined extremal point, the central charge is identically 0. Clearly this is inconsistent. This is in fact the same problem one would encounter if one naively specialises the general $\mathcal{N}=(0,2)$ twist of 4d $\mathcal{N}=4$ SYM in \cite{Benini:2013cda} to a $\mathcal{N}=(2,2)$ twist in the constant curvature Riemann surface case. The problem arises because $c$-extremization mixes the holomorphic (right-moving) and anti-holomorphic (left-moving) R-symmetries. We leave recovering the supergravity result from a field theory computation to the future.

\section*{Acknowledgments}
C. Couzens is supported by the Netherlands Organisation for Scientific research (NWO) under the VICI grant 680-47-602.
N. Macpherson is supported by the Spanish government grant PGC2018-096894-B100. A. Passias is supported by the Hellenic Foundation for Research and Innovation (H.F.R.I.) under the ``First Call for H.F.R.I.
Research Projects to support Faculty members and Researchers and
the procurement of high-cost research equipment grant'' (MIS 1857, Project Number: 16519), and by the LabEx ENS-ICFP: ANR-10-LABX-0010/ANR-10-IDEX-0001-02 PSL*.
\appendix

\section{Reducing the conditions}\label{sec:derivation}

We have seen that the gravity solution is fixed by determining a four-dimensional metric satisfying the torsion conditions \eqref{Jpsi}-\eqref{Omegad4}. Our goal in this section is to obtain the reduced conditions after substituting in an ansatz for the four-dimensional base. We will take the most general 4d metric admitting a single U$(1)$ isometry as used in \cite{Bah:2013qya} to study AdS$_5$ solutions in M-theory. Explicitly, the metric is
\be
\dd s^2_4 = \me^{2 A}(\dd x_1^2 +\dd x_2^2)+ \me^{2 B} \left[(\dd \theta +\me^{C} V^{R})^2+ \me^{2 C} (\dd \phi+ V^{I})^2\right]\, ,
\ee
where $V^{I}$ and $V^{R}$ have legs only along the Riemann surface parametrised by the coordinates $(x_1,\,x_2)$. The vector $\partial_{\phi}$ is taken to be a Killing vector and thus the three scalars $A,B,C$ are independent of $\phi$, though they may depend on the other three coordinates and $y$. We will use conventions in which the Hodge dual on the Riemann surface satisfies
\be
\star_2 \dd x_1=- \dd x_2\, ,\quad \star_2 \dd x_2= \dd x_1\, ,
\ee
and the volume form is given by
\be
\dd \vol_2=\dd x_1\wedge \dd x_2\, .
\ee
In addition, let us define the one-forms 
\be
\eta_{\theta}\equiv\dd \theta + \me^{C} V^R\, ,\quad \eta_{\phi}\equiv \dd \phi+ V^{I}\, ,
\ee
and the twisted exterior derivative on the Riemann surface
\be
\hat{\dd}_2\equiv  \dd_2-\me^C V^R \partial_{\theta}\, .
\ee
With the above metric ansatz the SU$(2)$-structure forms are 
\be
J=\me^{2 A} \dd \vol_2 +\me^{2B+C} \eta_\theta\wedge \eta_\phi\, ,\quad \Omega= \me^{A+B}\me^{-\ii \psi_1} (\dd x_1+\ii \dd x_2) \wedge (\eta_{\theta}+\ii \me^C \eta_{\phi})\, .\label{JOmansatz}
\ee

\subsection{Reducing the conditions }

Let us reduce the torsion conditions \eqref{Jpsi}-\eqref{Omegad4} on the ansatz above. We begin with the conditions for the holomorphic volume form before moving on to the K\"ahler form.


\subsection*{Reducing \eqref{Omegay}: $\partial_{y}\Omega$ equation}

From equation \eqref{Omegay} we obtain the conditions
\begin{align}
\partial_yC&=0\, ,\\
\star_2 V^R+V^I&=V^0\, ,\quad \partial_{y} V^0=0\, ,\label{V0def}\\
y \partial_y \log \Big(\me^{2(A+B)} \cos^2 \zeta\Big)&= - \tan^2 \zeta\, .
\end{align}
Since the first condition implies that the scalar $C$ is independent of the $y$ coordinate it follows that it may be removed by a change of coordinates. We will therefore set $C=0$ in the following. Moreover we may solve the final condition by introducing the scalar $\Lambda$:
\be\label{Lambdadef}
\cos^2 \zeta= - \frac{1}{y \partial_y \Lambda}\, ,\quad \me^{2(A+B)} =-y^2 \partial_{y} \me^{\Lambda}\, .
\ee


\subsection*{Reducing \eqref{Omegad4}: $\dd_4 \Omega$ equation}

Next consider \eqref{Omegad4}. Once the dust settles we obtain two conditions
\begin{align}
0&=\partial_{\theta} V^0\, ,\\
\sigma&=-  \frac{1}{2}\star_2 \hat{\dd}_2 \log \Big( \me^{2(A+B)}\cos^2\zeta\Big) +\frac{1}{2}\partial_{\theta} \log\Big(\me^{2(A+B)}\cos^2\zeta\Big) \eta_{\phi}\,,
\end{align}
the latter of which implies that $\sigma$ is determined by the function $\Lambda$ introduced in the previous section, 
\be
\sigma=-q \dd \phi -\frac{1}{2} \star_2 \hat{\dd}_2 \Lambda +\frac{1}{2} \partial_{\theta} \Lambda \eta_{\phi}\, .
\ee
This exhausts the non-trivial conditions arising from the holomorphic volume form, and we turn our attention to the K\"ahler form. 


\subsection*{Reducing \eqref{Jy}: $\partial_{y} J$ equation}

The conditions resulting from imposing \eqref{Jy} on the above ansatz are
\begin{align}
4 \partial_{y} \me^{2B}&= \partial_{\theta}^2 \Lambda\, ,\label{yJ1}\\
4 \me^{2 B}\partial_{y} V^{R}&= \hat{\dd}_{2}\partial_{\theta} \Lambda\, ,\label{yJ2}\\
4 \partial_{y}\me^{2 A} \dd \vol_{2}&= \partial_{\theta}\Lambda \hat{\dd}_{2} V^{I}-\hat{\dd}_{2} \star_{2} \hat{\dd}_{2} \Lambda\, ,\label{yJ3}\\
4 \me^{2B}\partial_{y} V^{I}&= \partial_{\theta} \Lambda \partial_{\theta}V^{I} -\partial_{\theta} \star_{2} \hat{\dd}_{2} \Lambda\, .\label{yJ4}
\end{align}
Note that the final condition, \eqref{yJ4} is implied by \eqref{yJ2} above and equation \eqref{V0def}.


\subsection*{Reducing \eqref{Jd4}: $\dd_4 J$ equation}

Finally imposing that the base is K\"ahler for fixed $y$ is equivalent to the three conditions
\begin{align}
\partial_{\theta} \me^{2A} \dd \vol_{2}&= \me^{2B} \hat{\dd}_{2} V^{I}\, ,\label{d4J1}\\
\hat{\dd}_{2} \me^{2 B} &= \me^{2B}\partial_{\theta} V^{R}\, ,\label{d4J2}\\
\hat{\dd}_{2}V^{R}&=0\label{d4J3}\, .
\end{align}


\subsection{Simplifying the conditions}
Restricting to the independent equations we are left with six independent conditions to solve, namely \eqref{yJ1}-\eqref{yJ3} and \eqref{d4J1}-\eqref{d4J3} subject to the two constraints \eqref{V0def} and \eqref{Lambdadef}. 

First note that equation \eqref{d4J3} implies that the twisted differential operator $\hat{\dd}_{2}$ is nilpotent. Moreover, it follows that $V^{R}$ is locally exact with respect to this twisted exterior derivative and we may write it (locally) as
\be
V^{R}= \hat{\dd}_{2} \Gamma\, ,
\ee
for some function $\Gamma$. To proceed it is useful to make the change of coordinates \cite{Bah:2013qya}
\be
X^{i}(x)= x^{i}\, ,\quad -\Gamma(X, \Theta, y) =\theta\,. 
\ee
With this change of coordinates the derivatives are given by 
\begin{align}
\partial_{\theta}&=-\frac{1}{\partial_{\Theta} \Gamma} \partial_{\Theta}\, ,\quad  \partial_{y}\rightarrow \partial_{y}- \frac{\partial_{Y}\Gamma}{\partial_{\Theta}\Gamma}\partial_{\Theta}\, ,\quad \partial_{x}= \partial_{X}- \frac{\partial_{X}\Gamma}{\partial_{\Theta}\Gamma} \partial_{\Theta}\, ,
\end{align}
whilst the twisted differential becomes
\be
\hat{\dd}_{2}= \dd^X_{2}\equiv \dd X^{i}\wedge \partial_{X^{i}}\, .
\ee
With this change of coordinates (and dropping the superscript $X$ on $\dd^X_2$ from now on) the one-form $\eta_{\theta}$ takes the simple form
\be
\eta_{\theta}= - \partial_{\Theta} \Gamma \dd \Theta-\partial_{y} \Gamma \dd Y\, .
\ee
We can now solve \eqref{d4J2} as
\be
\me^{2B}= -\frac{G}{\partial_{\Theta}\Gamma}\, ,\quad \dd_{2} G=0\, ,
\ee
with $G$ a scalar function. Inserting this into \eqref{Lambdadef} fixes the warp factor of the Riemann surface to be
\be
\me^{2 A}= \frac{y^2 \me^{\Lambda} g}{G}\, ,
\ee
where we have defined
\be
g \equiv \partial_y \Lambda \partial_{\Theta} \Gamma- \partial_y\Gamma \partial_\Theta \Lambda\, .
\ee
Since $V^0$ is independent of both $\theta$ and $y$ it follows  that for a suitable $X$ dependent gauge choice for $\Gamma$ we may, without loss of generality set $V^0=0$. It then follows that \eqref{d4J1} is equivalent to 
\be\label{boxGamma}
G \Big(\partial_{X_1}^2+\partial_{X_2}^2\Big) \Gamma= y^2 \partial_{\Theta} \bigg( \frac{\me^{\Lambda} g}{G}\bigg)\, .
\ee
Next consider equations \eqref{yJ1} and \eqref{yJ2}. They are equivalent to
\begin{align}
\partial_{y} G &= \partial_{\Theta} \widetilde{G}\, ,\\
\dd_2 \widetilde{G}&=0\, ,
\end{align}
where we have defined the function
\be
\widetilde{G}\equiv \frac{1}{4 \partial_{\Theta}\Gamma}\Big( 4 G \partial_y \Gamma -\partial_{\Theta} \Lambda\Big)\, . 
\ee
The final equation to solve is \eqref{yJ3} which implies
\be\label{boxLambda}
(G \partial_y -\widetilde{G}\partial_\Theta)\me^{2 A}=   \frac{G}{4} (\partial_{X_1}^2+\partial_{X_2}^2)\Lambda\, .
\ee
Having further reduced the conditions in terms of scalar potentials let us study the 
structure of the metric. First we define 
\be
h \equiv \partial_y \tilde{\Lambda}\partial_{\Theta} \Gamma-\partial_{\Theta} \tilde{\Lambda}\partial_{y} \Gamma\, ,\quad \text{with} \quad \tilde{\Lambda}=\Lambda+ \log y\, ,
\ee
then the warp factor is given by 
\be
\cos^2 \zeta=-\frac{\partial_{\Theta} \Gamma}{y g}\, ,\quad \text{and}\quad \sin^2\zeta = \frac{h}{ g}\, .
\ee
The ten-dimensional metric becomes
\be
\frac{1}{L^2}\dd s^2=\sqrt{\frac{y g}{h}}\bigg[ \dd s^2(\text{AdS}_3)+ \frac{h}{g} \dd \psi_2^2+\me^{-4 \Delta} \me^{2A} (\dd X_1^2+\dd X_2^2)+\me^{-4 \Delta} \dd s^2(\mathcal{M}_4)\bigg]\, ,
\ee
where
\be
\dd s^2(\mathcal{M}_4)= - \bigg[  \frac{g}{4 \partial_{\Theta} \Gamma}\dd y^2+\frac{\partial_{\Theta}\Gamma}{h}\Big( \eta_{\Theta}+ \frac{\partial_{\Theta}\Lambda}{\partial_{\Theta} \Gamma} \eta_{y}\Big)^2  +\partial_{\Theta} \Gamma G\Big(\dd \Theta +\frac{\partial_{y} \Gamma}{\partial_{\Theta} \Gamma} \dd y\Big)^2 +\frac{4 G}{\partial_{\Theta} \Gamma} \eta_{y}^2  \bigg]
\ee
and we have defined 
\be
\eta_{\Theta}\equiv\dd \psi_{1}-\frac{1}{2} \star_2 \dd_2 \Lambda\, ,\quad
\eta_{y}\equiv -\frac{1}{2}(\dd \phi-\star_2 \dd_2 \Gamma)\,,
\ee
where the reason for this labelling of the one-forms will become apparent soon. If we make the change of coordinates specified by
\be
\dd \bar{\theta}= G \dd \Theta+ \widetilde{G} \dd y\, ,
\ee
under which the partial derivatives transform as
\be
\partial_{\Theta}=G \partial_{\bar{\theta}}\, ,\quad \partial_{y}\rightarrow  \partial_{y}+ \widetilde{G} \partial_{\bar{\theta}}\, ,
\ee
the functions $G$ and $\widetilde{G}$ are eliminated from both the metric and the remaining conditions to solve. We may therefore set $G$ to any non-zero constant, set $\widetilde{G}=0$ and revert back to the previous set of coordinates. We fix $G=\frac{1}{4}$ so that the condition from setting $\widetilde{G}=0$ implies the integrability condition
\be
\partial_y \Gamma= \partial_{\Theta} \Lambda\, .
\ee
We may solve this without loss of generality by introducing a potential $D$ such that
\be
\Gamma= \partial_{\Theta} D\, \quad \Lambda= \partial_y D\, .
\ee
The two remaining conditions \eqref{boxGamma}  and \eqref{boxLambda} then imply
\be
\square D -16 g y^2 \me^{\partial_y D}=f(X_1,X_2)\, .
\ee
However, by suitably redefining the potential $D$ we can set the function $f(X_1,X_2)$ to zero without loss of generality, so we do so. We must therefore solve the Mong\'e--Ampere like equation
\be
\square D=16 y^2\Big(\partial_y^2 D \partial_{\Theta}^2 D- (\partial_y \partial_\Theta D)^2\Big) \me^{\partial_y D}\, .
\ee
The final metric is 
\begin{align}
\frac{1}{L^2}\dd s^2=&\, \sqrt{\frac{y g}{h}}\bigg[ \dd s^2(\text{AdS}_3)+ \frac{h}{g} \dd \psi_2^2 + \frac{ h}{y g} \me^{2 A} (\dd X_1^2 +\dd X_2^2) + \frac{h}{y g} \dd s^2(\mathcal{M}_4)\bigg]\, ,\\[2mm]
\dd s^2(\mathcal{M}_4)=&\,  \frac{1}{4}g_{ij} \dd u^{i} \dd u^{j}+ h^{ij}\eta_{i}\eta_{j}\, ,\\[2mm]
g_{ij}\equiv&\, -\partial_{i}\partial_{j} D\, ,\\[2mm]
h_{ij}\equiv&\, -\partial_{i}\partial_{j} \tilde{D}\, ,
\end{align}
where 
\be
u^{i}=\{ y,\Theta\}\, ,\quad \eta_1=\eta_{\Theta}\, ,\quad \eta_2=-\eta_y\, ,
\ee
and 
\be
\tilde{D}\equiv D+y( \log y-1)\,,\quad e^{2A}= 4  y^2e^{\partial_{y}D} g.
\ee
Note that the functions $g$ and $h$ are now the determinants of the matrices $g_{ij}$ and $h_{ij}$ respectively. \\
~~\\
The flux is given by
\be
F^{(2)}=-2 J +y \dd\Big(\eta_\Theta -\frac{\partial_{\Theta}\Lambda}{\partial_\Theta \Gamma} \eta_y\Big)  -\dd \bigg(\frac{y g}{h}\Big(\eta_\Theta -\frac{\partial_{\Theta}\Lambda}{\partial_\Theta \Gamma} \eta_y\Big)\bigg)\, ,
\ee
with $J$ as given in \eqref{JOmansatz}.

\bibliographystyle{JHEP}
\bibliography{ADSCFT}

\providecommand{\href}[2]{#2}\begingroup\raggedright\begin{thebibliography}{10}

\bibitem{Boido:2021szx}
A.~Boido, J.~M.~P. Ipi\~na and J.~Sparks, \emph{{Twisted D3-brane and M5-brane
  compactifications from multi-charge spindles}},
  \href{http://dx.doi.org/10.1007/JHEP07(2021)222}{\emph{JHEP} {\bf 07} (2021)
  222}, [\href{https://arxiv.org/abs/2104.13287}{{\tt 2104.13287}}].

\bibitem{Maldacena:2000mw}
J.~M. Maldacena and C.~Nunez, \emph{{Supergravity description of field theories
  on curved manifolds and a no go theorem}},
  \href{http://dx.doi.org/10.1142/S0217751X01003937}{\emph{Int. J. Mod. Phys.
  A} {\bf 16} (2001) 822--855},
  [\href{https://arxiv.org/abs/hep-th/0007018}{{\tt hep-th/0007018}}].

\bibitem{Gaiotto:2009gz}
D.~Gaiotto and J.~Maldacena, \emph{{The Gravity duals of N=2 superconformal
  field theories}},
  \href{http://dx.doi.org/10.1007/JHEP10(2012)189}{\emph{JHEP} {\bf 10} (2012)
  189}, [\href{https://arxiv.org/abs/0904.4466}{{\tt 0904.4466}}].

\bibitem{Bah:2015fwa}
I.~Bah, \emph{{AdS5 solutions from M5-branes on Riemann surface and D6-branes
  sources}}, \href{http://dx.doi.org/10.1007/JHEP09(2015)163}{\emph{JHEP} {\bf
  09} (2015) 163}, [\href{https://arxiv.org/abs/1501.06072}{{\tt 1501.06072}}].

\bibitem{Bah:2011je}
I.~Bah and B.~Wecht, \emph{{New N=1 Superconformal Field Theories In Four
  Dimensions}}, \href{http://dx.doi.org/10.1007/JHEP07(2013)107}{\emph{JHEP}
  {\bf 07} (2013) 107}, [\href{https://arxiv.org/abs/1111.3402}{{\tt
  1111.3402}}].

\bibitem{Bah:2012dg}
I.~Bah, C.~Beem, N.~Bobev and B.~Wecht, \emph{{Four-Dimensional SCFTs from
  M5-Branes}}, \href{http://dx.doi.org/10.1007/JHEP06(2012)005}{\emph{JHEP}
  {\bf 06} (2012) 005}, [\href{https://arxiv.org/abs/1203.0303}{{\tt
  1203.0303}}].

\bibitem{Bah:2013qya}
I.~Bah, \emph{{Quarter-BPS $AdS_{5}$ solutions in M-theory with a $T^{2}$
  bundle over a Riemann surface}},
  \href{http://dx.doi.org/10.1007/JHEP08(2013)137}{\emph{JHEP} {\bf 08} (2013)
  137}, [\href{https://arxiv.org/abs/1304.4954}{{\tt 1304.4954}}].

\bibitem{Bah:2021mzw}
I.~Bah, F.~Bonetti, R.~Minasian and E.~Nardoni, \emph{{Holographic Duals of
  Argyres-Douglas Theories}},
  \href{http://dx.doi.org/10.1103/PhysRevLett.127.211601}{\emph{Phys. Rev.
  Lett.} {\bf 127} (2021) 211601},
  [\href{https://arxiv.org/abs/2105.11567}{{\tt 2105.11567}}].

\bibitem{Bah:2021hei}
I.~Bah, F.~Bonetti, R.~Minasian and E.~Nardoni, \emph{{M5-brane sources,
  holography, and Argyres-Douglas theories}},
  \href{http://dx.doi.org/10.1007/JHEP11(2021)140}{\emph{JHEP} {\bf 11} (2021)
  140}, [\href{https://arxiv.org/abs/2106.01322}{{\tt 2106.01322}}].

\bibitem{Bobev:2019zmz}
N.~Bobev and P.~M. Crichigno, \emph{{Universal spinning black holes and
  theories of class $ \mathcal{R} $}},
  \href{http://dx.doi.org/10.1007/JHEP12(2019)054}{\emph{JHEP} {\bf 12} (2019)
  054}, [\href{https://arxiv.org/abs/1909.05873}{{\tt 1909.05873}}].

\bibitem{Bah:2018lyv}
I.~Bah, A.~Passias and P.~Weck, \emph{{Holographic duals of five-dimensional
  SCFTs on a Riemann surface}},
  \href{http://dx.doi.org/10.1007/JHEP01(2019)058}{\emph{JHEP} {\bf 01} (2019)
  058}, [\href{https://arxiv.org/abs/1807.06031}{{\tt 1807.06031}}].

\bibitem{Bobev:2019ore}
N.~Bobev, P.~Bomans and F.~F. Gautason, \emph{{Wrapped Branes and Punctured
  Horizons}}, \href{http://dx.doi.org/10.1007/JHEP06(2020)011}{\emph{JHEP} {\bf
  06} (2020) 011}, [\href{https://arxiv.org/abs/1912.04779}{{\tt 1912.04779}}].

\bibitem{Ferrero:2020laf}
P.~Ferrero, J.~P. Gauntlett, J.~M. P\'erez Ipi\~na, D.~Martelli and J.~Sparks,
  \emph{{D3-Branes Wrapped on a Spindle}},
  \href{http://dx.doi.org/10.1103/PhysRevLett.126.111601}{\emph{Phys. Rev.
  Lett.} {\bf 126} (2021) 111601},
  [\href{https://arxiv.org/abs/2011.10579}{{\tt 2011.10579}}].

\bibitem{Ferrero:2021wvk}
P.~Ferrero, J.~P. Gauntlett, D.~Martelli and J.~Sparks, \emph{{M5-branes
  wrapped on a spindle}},
  \href{http://dx.doi.org/10.1007/JHEP11(2021)002}{\emph{JHEP} {\bf 11} (2021)
  002}, [\href{https://arxiv.org/abs/2105.13344}{{\tt 2105.13344}}].

\bibitem{Hosseini:2021fge}
S.~M. Hosseini, K.~Hristov and A.~Zaffaroni, \emph{{Rotating multi-charge
  spindles and their microstates}},
  \href{http://dx.doi.org/10.1007/JHEP07(2021)182}{\emph{JHEP} {\bf 07} (2021)
  182}, [\href{https://arxiv.org/abs/2104.11249}{{\tt 2104.11249}}].

\bibitem{Belin:2020nmp}
A.~Belin, N.~Benjamin, A.~Castro, S.~M. Harrison and C.~A. Keller,
  \emph{{$\mathcal{N}=2$ Minimal Models: A Holographic Needle in a Symmetric
  Orbifold Haystack}},
  \href{http://dx.doi.org/10.21468/SciPostPhys.8.6.084}{\emph{SciPost Phys.}
  {\bf 8} (2020) 084}, [\href{https://arxiv.org/abs/2002.07819}{{\tt
  2002.07819}}].

\bibitem{Belin:2019jqz}
A.~Belin, A.~Castro, C.~A. Keller and B.~J. M\"uhlmann, \emph{{Siegel
  Paramodular Forms from Exponential Lifts: Slow versus Fast Growth}},
  \href{https://arxiv.org/abs/1910.05353}{{\tt 1910.05353}}.

\bibitem{Belin:2019rba}
A.~Belin, A.~Castro, C.~A. Keller and B.~M\"uhlmann, \emph{{The Holographic
  Landscape of Symmetric Product Orbifolds}},
  \href{http://dx.doi.org/10.1007/JHEP01(2020)111}{\emph{JHEP} {\bf 01} (2020)
  111}, [\href{https://arxiv.org/abs/1910.05342}{{\tt 1910.05342}}].

\bibitem{Azzola:2018sld}
M.~Azzola, D.~Klemm and M.~Rabbiosi, \emph{{AdS$_5$ black strings in the stu
  model of FI-gauged $N=2$ supergravity}},
  \href{http://dx.doi.org/10.1007/JHEP10(2018)080}{\emph{JHEP} {\bf 10} (2018)
  080}, [\href{https://arxiv.org/abs/1803.03570}{{\tt 1803.03570}}].

\bibitem{Couzens:2017nnr}
C.~Couzens, D.~Martelli and S.~Schafer-Nameki, \emph{{F-theory and
  AdS$_{3}$/CFT$_{2}$ (2, 0)}},
  \href{http://dx.doi.org/10.1007/JHEP06(2018)008}{\emph{JHEP} {\bf 06} (2018)
  008}, [\href{https://arxiv.org/abs/1712.07631}{{\tt 1712.07631}}].

\bibitem{Martelli:2003ki}
D.~Martelli and J.~Sparks, \emph{{G structures, fluxes and calibrations in M
  theory}}, \href{http://dx.doi.org/10.1103/PhysRevD.68.085014}{\emph{Phys.
  Rev.} {\bf D68} (2003) 085014},
  [\href{https://arxiv.org/abs/hep-th/0306225}{{\tt hep-th/0306225}}].

\bibitem{Tsimpis:2005kj}
D.~Tsimpis, \emph{{M-theory on eight-manifolds revisited: N=1 supersymmetry and
  generalized spin(7) structures}},
  \href{http://dx.doi.org/10.1088/1126-6708/2006/04/027}{\emph{JHEP} {\bf 04}
  (2006) 027}, [\href{https://arxiv.org/abs/hep-th/0511047}{{\tt
  hep-th/0511047}}].

\bibitem{Kim:2005ez}
N.~Kim, \emph{{AdS(3) solutions of IIB supergravity from D3-branes}},
  \href{http://dx.doi.org/10.1088/1126-6708/2006/01/094}{\emph{JHEP} {\bf 01}
  (2006) 094}, [\href{https://arxiv.org/abs/hep-th/0511029}{{\tt
  hep-th/0511029}}].

\bibitem{Kim:2007hv}
H.~Kim, K.~K. Kim and N.~Kim, \emph{{1/4-BPS M-theory bubbles with SO(3) x
  SO(4) symmetry}},
  \href{http://dx.doi.org/10.1088/1126-6708/2007/08/050}{\emph{JHEP} {\bf 08}
  (2007) 050}, [\href{https://arxiv.org/abs/0706.2042}{{\tt 0706.2042}}].

\bibitem{Figueras:2007cn}
P.~Figueras, O.~A. Mac~Conamhna and E.~O~Colgain, \emph{{Global geometry of the
  supersymmetric AdS(3)/CFT(2) correspondence in M-theory}},
  \href{http://dx.doi.org/10.1103/PhysRevD.76.046007}{\emph{Phys. Rev. D} {\bf
  76} (2007) 046007}, [\href{https://arxiv.org/abs/hep-th/0703275}{{\tt
  hep-th/0703275}}].

\bibitem{Donos:2008hd}
A.~Donos, J.~P. Gauntlett and J.~Sparks, \emph{{AdS(3) x (S**3 x S**3 x S**1)
  Solutions of Type IIB String Theory}},
  \href{http://dx.doi.org/10.1088/0264-9381/26/6/065009}{\emph{Class. Quant.
  Grav.} {\bf 26} (2009) 065009}, [\href{https://arxiv.org/abs/0810.1379}{{\tt
  0810.1379}}].

\bibitem{Colgain:2010wb}
E.~O~Colgain, J.-B. Wu and H.~Yavartanoo, \emph{{Supersymmetric AdS3 X S2
  M-theory geometries with fluxes}},
  \href{http://dx.doi.org/10.1007/JHEP08(2010)114}{\emph{JHEP} {\bf 08} (2010)
  114}, [\href{https://arxiv.org/abs/1005.4527}{{\tt 1005.4527}}].

\bibitem{DHoker:2008lup}
E.~D'Hoker, J.~Estes, M.~Gutperle and D.~Krym, \emph{{Exact Half-BPS Flux
  Solutions in M-theory. I: Local Solutions}},
  \href{http://dx.doi.org/10.1088/1126-6708/2008/08/028}{\emph{JHEP} {\bf 08}
  (2008) 028}, [\href{https://arxiv.org/abs/0806.0605}{{\tt 0806.0605}}].

\bibitem{Estes:2012vm}
J.~Estes, R.~Feldman and D.~Krym, \emph{{Exact half-BPS flux solutions in $M$
  theory with D(2,1;$c^\prime$;0)$^2$ symmetry: Local solutions}},
  \href{http://dx.doi.org/10.1103/PhysRevD.87.046008}{\emph{Phys. Rev.} {\bf
  D87} (2013) 046008}, [\href{https://arxiv.org/abs/1209.1845}{{\tt
  1209.1845}}].

\bibitem{Bachas:2013vza}
C.~Bachas, E.~D'Hoker, J.~Estes and D.~Krym, \emph{{M-theory Solutions
  Invariant under $D(2,1;\gamma) \oplus D(2,1;\gamma)$}},
  \href{http://dx.doi.org/10.1002/prop.201300039}{\emph{Fortsch. Phys.} {\bf
  62} (2014) 207--254}, [\href{https://arxiv.org/abs/1312.5477}{{\tt
  1312.5477}}].

\bibitem{Jeong:2014iva}
J.~Jeong, E.~O~Colgain and K.~Yoshida, \emph{{SUSY properties of warped
  $AdS_3$}}, \href{http://dx.doi.org/10.1007/JHEP06(2014)036}{\emph{JHEP} {\bf
  06} (2014) 036}, [\href{https://arxiv.org/abs/1402.3807}{{\tt 1402.3807}}].

\bibitem{Lozano:2015bra}
Y.~Lozano, N.~T. Macpherson, J.~Montero and E.~O. Colgain, \emph{{New $AdS_3
  \times S^2$ T-duals with $ \mathcal{N}=\left(0,4\right) $ supersymmetry}},
  \href{http://dx.doi.org/10.1007/JHEP08(2015)121}{\emph{JHEP} {\bf 08} (2015)
  121}, [\href{https://arxiv.org/abs/1507.02659}{{\tt 1507.02659}}].

\bibitem{Kelekci:2016uqv}
O.~Kelekci, Y.~Lozano, J.~Montero, E.~Colgain and M.~Park, \emph{{Large
  superconformal near-horizons from M-theory}},
  \href{http://dx.doi.org/10.1103/PhysRevD.93.086010}{\emph{Phys. Rev. D} {\bf
  93} (2016) 086010}, [\href{https://arxiv.org/abs/1602.02802}{{\tt
  1602.02802}}].

\bibitem{Couzens:2017way}
C.~Couzens, C.~Lawrie, D.~Martelli, S.~Schafer-Nameki and J.-M. Wong,
  \emph{{F-theory and AdS$_{3}$/CFT$_{2}$}},
  \href{http://dx.doi.org/10.1007/JHEP08(2017)043}{\emph{JHEP} {\bf 08} (2017)
  043}, [\href{https://arxiv.org/abs/1705.04679}{{\tt 1705.04679}}].

\bibitem{Eberhardt:2017uup}
L.~Eberhardt, \emph{{Supersymmetric AdS$_{3}$ supergravity backgrounds and
  holography}}, \href{http://dx.doi.org/10.1007/JHEP02(2018)087}{\emph{JHEP}
  {\bf 02} (2018) 087}, [\href{https://arxiv.org/abs/1710.09826}{{\tt
  1710.09826}}].

\bibitem{Dibitetto:2018iar}
G.~Dibitetto and N.~Petri, \emph{{Surface defects in the D4 $-$ D8 brane
  system}}, \href{http://dx.doi.org/10.1007/JHEP01(2019)193}{\emph{JHEP} {\bf
  01} (2019) 193}, [\href{https://arxiv.org/abs/1807.07768}{{\tt 1807.07768}}].

\bibitem{Dibitetto:2018ftj}
G.~Dibitetto, G.~Lo~Monaco, A.~Passias, N.~Petri and A.~Tomasiello,
  \emph{{AdS$_3$ Solutions with Exceptional Supersymmetry}},
  \href{http://dx.doi.org/10.1002/prop.201800060}{\emph{Fortsch. Phys.} {\bf
  66} (2018) 1800060}, [\href{https://arxiv.org/abs/1807.06602}{{\tt
  1807.06602}}].

\bibitem{Macpherson:2018mif}
N.~T. Macpherson, \emph{{Type II solutions on AdS$_{3} \times$ S$^{3} \times$
  S$^{3}$ with large superconformal symmetry}},
  \href{http://dx.doi.org/10.1007/JHEP05(2019)089}{\emph{JHEP} {\bf 05} (2019)
  089}, [\href{https://arxiv.org/abs/1812.10172}{{\tt 1812.10172}}].

\bibitem{Legramandi:2019xqd}
A.~Legramandi and N.~T. Macpherson, \emph{{AdS$_3$ solutions with from
  $\mathcal{N}=(3,0)$ from S$^3\times$S$^3$ fibrations}},
  \href{http://dx.doi.org/10.1002/prop.202000014}{\emph{Fortsch. Phys.} {\bf
  68} (2020) 2000014}, [\href{https://arxiv.org/abs/1912.10509}{{\tt
  1912.10509}}].

\bibitem{Lozano:2019emq}
Y.~Lozano, N.~T. Macpherson, C.~Nunez and A.~Ramirez, \emph{{AdS$_3$ solutions
  in Massive IIA with small $\mathcal{N}=(4,0)$ supersymmetry}},
  \href{http://dx.doi.org/10.1007/JHEP01(2020)129}{\emph{JHEP} {\bf 01} (2020)
  129}, [\href{https://arxiv.org/abs/1908.09851}{{\tt 1908.09851}}].

\bibitem{Lozano:2019jza}
Y.~Lozano, N.~T. Macpherson, C.~Nunez and A.~Ramirez, \emph{{1/4 BPS solutions
  and the AdS$_3$/CFT$_2$ correspondence}},
  \href{http://dx.doi.org/10.1103/PhysRevD.101.026014}{\emph{Phys. Rev.} {\bf
  D101} (2020) 026014}, [\href{https://arxiv.org/abs/1909.09636}{{\tt
  1909.09636}}].

\bibitem{Lozano:2019zvg}
Y.~Lozano, N.~T. Macpherson, C.~Nunez and A.~Ramirez, \emph{{Two dimensional
  ${\cal N}=(0,4)$ quivers dual to AdS$_3$ solutions in massive IIA}},
  \href{http://dx.doi.org/10.1007/JHEP01(2020)140}{\emph{JHEP} {\bf 01} (2020)
  140}, [\href{https://arxiv.org/abs/1909.10510}{{\tt 1909.10510}}].

\bibitem{Lozano:2019ywa}
Y.~Lozano, N.~T. Macpherson, C.~Nunez and A.~Ramirez, \emph{{AdS$_3$ solutions
  in massive IIA, defect CFTs and T-duality}},
  \href{http://dx.doi.org/10.1007/JHEP12(2019)013}{\emph{JHEP} {\bf 12} (2019)
  013}, [\href{https://arxiv.org/abs/1909.11669}{{\tt 1909.11669}}].

\bibitem{Couzens:2019mkh}
C.~Couzens, H.~het Lam and K.~Mayer, \emph{{Twisted $ \mathcal{N} $ = 1 SCFTs
  and their AdS$_{3}$ duals}},
  \href{http://dx.doi.org/10.1007/JHEP03(2020)032}{\emph{JHEP} {\bf 03} (2020)
  032}, [\href{https://arxiv.org/abs/1912.07605}{{\tt 1912.07605}}].

\bibitem{Couzens:2019iog}
C.~Couzens, \emph{{$ \mathcal{N} $ = (0, 2) AdS$_{3}$ solutions of type IIB and
  F-theory with generic fluxes}},
  \href{http://dx.doi.org/10.1007/JHEP04(2021)038}{\emph{JHEP} {\bf 04} (2021)
  038}, [\href{https://arxiv.org/abs/1911.04439}{{\tt 1911.04439}}].

\bibitem{Passias:2019rga}
A.~Passias and D.~Prins, \emph{{On AdS$_3$ solutions of Type IIB}},
  \href{http://dx.doi.org/10.1007/JHEP05(2020)048}{\emph{JHEP} {\bf 05} (2020)
  048}, [\href{https://arxiv.org/abs/1910.06326}{{\tt 1910.06326}}].

\bibitem{Filippas:2019ihy}
K.~Filippas, \emph{{Non-integrability on AdS$_{3}$ supergravity backgrounds}},
  \href{http://dx.doi.org/10.1007/JHEP02(2020)027}{\emph{JHEP} {\bf 02} (2020)
  027}, [\href{https://arxiv.org/abs/1910.12981}{{\tt 1910.12981}}].

\bibitem{Speziali:2019uzn}
S.~Speziali, \emph{{Spin 2 fluctuations in 1/4 BPS AdS$_3$/CFT$_2$}},
  \href{http://dx.doi.org/10.1007/JHEP03(2020)079}{\emph{JHEP} {\bf 03} (2020)
  079}, [\href{https://arxiv.org/abs/1910.14390}{{\tt 1910.14390}}].

\bibitem{Lozano:2020bxo}
Y.~Lozano, C.~Nunez, A.~Ramirez and S.~Speziali, \emph{{$M$-strings and AdS$_3$
  solutions to M-theory with small $\mathcal{N}=(0,4)$ supersymmetry}},
  \href{http://dx.doi.org/10.1007/JHEP08(2020)118}{\emph{JHEP} {\bf 08} (2020)
  118}, [\href{https://arxiv.org/abs/2005.06561}{{\tt 2005.06561}}].

\bibitem{Farakos:2020phe}
F.~Farakos, G.~Tringas and T.~Van~Riet, \emph{{No-scale and scale-separated
  flux vacua from IIA on G2 orientifolds}},
  \href{http://dx.doi.org/10.1140/epjc/s10052-020-8247-5}{\emph{Eur. Phys. J.
  C} {\bf 80} (2020) 659}, [\href{https://arxiv.org/abs/2005.05246}{{\tt
  2005.05246}}].

\bibitem{Couzens:2020aat}
C.~Couzens, H.~het Lam, K.~Mayer and S.~Vandoren, \emph{{Anomalies of (0,4)
  SCFTs from F-theory}},
  \href{http://dx.doi.org/10.1007/JHEP08(2020)060}{\emph{JHEP} {\bf 08} (2020)
  060}, [\href{https://arxiv.org/abs/2006.07380}{{\tt 2006.07380}}].

\bibitem{Rigatos:2020igd}
K.~S. Rigatos, \emph{{Non-integrability in AdS$_{3}$ vacua}},
  \href{http://dx.doi.org/10.1007/JHEP02(2021)032}{\emph{JHEP} {\bf 02} (2021)
  032}, [\href{https://arxiv.org/abs/2011.08224}{{\tt 2011.08224}}].

\bibitem{Faedo:2020nol}
F.~Faedo, Y.~Lozano and N.~Petri, \emph{{Searching for surface defect CFTs
  within AdS$_3$}},
  \href{http://dx.doi.org/10.1007/JHEP11(2020)052}{\emph{JHEP} {\bf 11} (2020)
  052}, [\href{https://arxiv.org/abs/2007.16167}{{\tt 2007.16167}}].

\bibitem{Dibitetto:2020bsh}
G.~Dibitetto and N.~Petri, \emph{{AdS$_{3}$ from M-branes at conical
  singularities}}, \href{http://dx.doi.org/10.1007/JHEP01(2021)129}{\emph{JHEP}
  {\bf 01} (2021) 129}, [\href{https://arxiv.org/abs/2010.12323}{{\tt
  2010.12323}}].

\bibitem{Filippas:2020qku}
K.~Filippas, \emph{{Holography for 2D $\mathcal{N}=(0,4)$ quantum field
  theory}}, \href{http://dx.doi.org/10.1103/PhysRevD.103.086003}{\emph{Phys.
  Rev. D} {\bf 103} (2021) 086003},
  [\href{https://arxiv.org/abs/2008.00314}{{\tt 2008.00314}}].

\bibitem{Passias:2020ubv}
A.~Passias and D.~Prins, \emph{{On supersymmetric AdS$_3$ solutions of Type
  II}}, \href{http://dx.doi.org/10.1007/JHEP08(2021)168}{\emph{JHEP} {\bf 08}
  (2021) 168}, [\href{https://arxiv.org/abs/2011.00008}{{\tt 2011.00008}}].

\bibitem{Faedo:2020lyw}
F.~Faedo, Y.~Lozano and N.~Petri, \emph{{New $\mathcal{N}=(0,4)$ AdS$_3$
  near-horizons in Type IIB}},
  \href{http://dx.doi.org/10.1007/JHEP04(2021)028}{\emph{JHEP} {\bf 04} (2021)
  028}, [\href{https://arxiv.org/abs/2012.07148}{{\tt 2012.07148}}].

\bibitem{Eloy:2020uix}
C.~Eloy, \emph{{Kaluza-Klein spectrometry for ${\rm AdS_{3}}$ vacua}},
  \href{http://dx.doi.org/10.21468/SciPostPhys.10.6.131}{\emph{SciPost Phys.}
  {\bf 10} (2021) 131}, [\href{https://arxiv.org/abs/2011.11658}{{\tt
  2011.11658}}].

\bibitem{Legramandi:2020txf}
A.~Legramandi, G.~Lo~Monaco and N.~T. Macpherson, \emph{{All
  $\mathcal{N}=(8,0)$ AdS$_3$ solutions in 10 and 11 dimensions}},
  \href{http://dx.doi.org/10.1007/JHEP05(2021)263}{\emph{JHEP} {\bf 05} (2021)
  263}, [\href{https://arxiv.org/abs/2012.10507}{{\tt 2012.10507}}].

\bibitem{Zacarias:2021pfz}
S.~Zacarias, \emph{{Marginal deformations of a class of AdS$_{3}$ $\mathcal{N}
  $ = (0, 4) holographic backgrounds}},
  \href{http://dx.doi.org/10.1007/JHEP06(2021)017}{\emph{JHEP} {\bf 06} (2021)
  017}, [\href{https://arxiv.org/abs/2102.05681}{{\tt 2102.05681}}].

\bibitem{Emelin:2021gzx}
M.~Emelin, F.~Farakos and G.~Tringas, \emph{{Three-dimensional flux vacua from
  IIB on co-calibrated G2 orientifolds}},
  \href{http://dx.doi.org/10.1140/epjc/s10052-021-09261-y}{\emph{Eur. Phys. J.
  C} {\bf 81} (2021) 456}, [\href{https://arxiv.org/abs/2103.03282}{{\tt
  2103.03282}}].

\bibitem{Couzens:2019wls}
C.~Couzens, H.~het Lam, K.~Mayer and S.~Vandoren, \emph{{Black Holes and (0,4)
  SCFTs from Type IIB on K3}},
  \href{http://dx.doi.org/10.1007/JHEP08(2019)043}{\emph{JHEP} {\bf 08} (2019)
  043}, [\href{https://arxiv.org/abs/1904.05361}{{\tt 1904.05361}}].

\bibitem{Passias:2018zlm}
A.~Passias, D.~Prins and A.~Tomasiello, \emph{{A massive class of $\mathcal{N}
  = 2$ AdS$_4$ IIA solutions}},
  \href{http://dx.doi.org/10.1007/JHEP10(2018)071}{\emph{JHEP} {\bf 10} (2018)
  071}, [\href{https://arxiv.org/abs/1805.03661}{{\tt 1805.03661}}].

\bibitem{Gauntlett:2004zh}
J.~P. Gauntlett, D.~Martelli, J.~Sparks and D.~Waldram, \emph{{Supersymmetric
  AdS(5) solutions of M theory}},
  \href{http://dx.doi.org/10.1088/0264-9381/21/18/005}{\emph{Class. Quant.
  Grav.} {\bf 21} (2004) 4335--4366},
  [\href{https://arxiv.org/abs/hep-th/0402153}{{\tt hep-th/0402153}}].

\bibitem{Bah:2015nva}
I.~Bah and V.~Stylianou, \emph{{Gravity duals of $ \mathcal{N}=\left(0,\
  2\right) $ SCFTs from M5-branes}},
  \href{http://dx.doi.org/10.1007/JHEP04(2019)050}{\emph{JHEP} {\bf 04} (2019)
  050}, [\href{https://arxiv.org/abs/1508.04135}{{\tt 1508.04135}}].

\bibitem{Bah:2019rgq}
I.~Bah, F.~Bonetti, R.~Minasian and E.~Nardoni, \emph{{Anomalies of QFTs from
  M-theory and Holography}},
  \href{http://dx.doi.org/10.1007/JHEP01(2020)125}{\emph{JHEP} {\bf 01} (2020)
  125}, [\href{https://arxiv.org/abs/1910.04166}{{\tt 1910.04166}}].

\bibitem{Lawrie:2016axq}
C.~Lawrie, S.~Schafer-Nameki and T.~Weigand, \emph{{Chiral 2d theories from N =
  4 SYM with varying coupling}},
  \href{http://dx.doi.org/10.1007/JHEP04(2017)111}{\emph{JHEP} {\bf 04} (2017)
  111}, [\href{https://arxiv.org/abs/1612.05640}{{\tt 1612.05640}}].

\bibitem{Benini:2013cda}
F.~Benini and N.~Bobev, \emph{{Two-dimensional SCFTs from wrapped branes and
  c-extremization}},
  \href{http://dx.doi.org/10.1007/JHEP06(2013)005}{\emph{JHEP} {\bf 06} (2013)
  005}, [\href{https://arxiv.org/abs/1302.4451}{{\tt 1302.4451}}].

\bibitem{Amariti:2017cyd}
A.~Amariti, L.~Cassia and S.~Penati, \emph{{Surveying 4d SCFTs twisted on
  Riemann surfaces}},
  \href{http://dx.doi.org/10.1007/JHEP06(2017)056}{\emph{JHEP} {\bf 06} (2017)
  056}, [\href{https://arxiv.org/abs/1703.08201}{{\tt 1703.08201}}].

\bibitem{Brown:1986tm}
J.~D. Brown and M.~Henneaux, \emph{Central charges in the canonical realization
  of asymptotic symmetries: An example from three dimensional gravity},
  \href{http://dx.doi.org/10.1007/BF01211590}{\emph{Communications in
  Mathematical Physics} {\bf 104} (1986) 207--226}.

\bibitem{Couzens:2018wnk}
C.~Couzens, J.~P. Gauntlett, D.~Martelli and J.~Sparks, \emph{{A geometric dual
  of $c$-extremization}},
  \href{http://dx.doi.org/10.1007/JHEP01(2019)212}{\emph{JHEP} {\bf 01} (2019)
  212}, [\href{https://arxiv.org/abs/1810.11026}{{\tt 1810.11026}}].

\bibitem{Gauntlett:2006ns}
J.~P. Gauntlett, N.~Kim and D.~Waldram, \emph{{Supersymmetric AdS(3), AdS(2)
  and Bubble Solutions}},
  \href{http://dx.doi.org/10.1088/1126-6708/2007/04/005}{\emph{JHEP} {\bf 04}
  (2007) 005}, [\href{https://arxiv.org/abs/hep-th/0612253}{{\tt
  hep-th/0612253}}].

\end{thebibliography}\endgroup

\end{document}